\documentclass{article}

\usepackage{microtype}
\usepackage{natbib}
\usepackage{graphicx}
\usepackage{subfigure}
\usepackage[T1]{fontenc}
\usepackage{subfiles}
\usepackage{amssymb}
\usepackage{amsmath}
\usepackage{amsthm}
\usepackage{mdframed}
\usepackage{mathtools}
\usepackage{enumitem}
\usepackage{dirtytalk}
\usepackage{bm}
\usepackage[dvipsnames]{xcolor}
\usepackage{hyperref}
\usepackage[capitalize,noabbrev]{cleveref}
\usepackage{xcolor}
\usepackage[ruled, vlined, linesnumbered]{algorithm2e}
\usepackage{bbm}
\usepackage{thm-restate}
\usepackage[a4paper, margin=1in]{geometry}
\theoremstyle{plain}
\newif\ifMyFlag
\MyFlagtrue

\newtheorem{theorem}{\textbf{Theorem}}[section]
\newtheorem{corollary}[theorem]{\textbf{Corollary}}

\newtheorem{definition}[theorem]{\textbf{Definition}}
\newtheorem{observation}[section]{\textbf{Observation}}

\newcommand{\N}{\mathbb{N}}
\newcommand{\R}{\mathbb{R}}
\newcommand{\Z}{\mathbb{Z}}
\newcommand{\bo}[1]{\mathcal{O} \big(#1\big)}
\newcommand{\norm}[2]{{\lVert #1 \rVert}_{#2}}
\newcommand{\ceil}[1]{\left \lceil #1 \right \rceil}

\newcommand{\set}[2]{\{ #1, \dotso, #2 \}}
\newcommand{\qt}[1]{\text{\say{#1}}}
\newcommand{\ILP}{\textsc{ILP Feasibility}\xspace}
\newcommand{\NUCS}{\textsc{Non-uniform Closest String}\xspace}
\newcommand{\CS}{\textsc{Closest String}\xspace}
\newcommand{\kCME}{\textsc{$k$-Center with Missing Entries}\xspace}
\newcommand{\kC}{\textsc{$k$-Center}\xspace}
\newcommand{\kMME}{\textsc{$k$-Means with Missing Entries}\xspace}

\newcommand{\fa}[1]{{\large \color{Plum} [Farehe: #1]}}

\newcommand{\A}{\bm{A}}
\newcommand{\B}{\bm{B}}
\newcommand{\M}{\bm{M}}
\newcommand{\x}{\bm{x}}
\newcommand{\vxc}{S}
\newcommand{\rvc}{R_{S}}
\newcommand{\cvc}{C_{S}}
\newcommand{\down}[1]{{#1}^{\scriptscriptstyle{\downarrow}}}
\newcommand{\hd}{d_H}

\newcommand{\figpath}{./figures}

\DeclareMathOperator{\cent}{cent}
\DeclareMathOperator{\parti}{part}
\DeclareMathOperator{\dist}{dist}
\DeclareMathOperator{\Dp}{D}
\DeclareMathOperator{\poly}{poly}
\DeclareMathOperator{\vc}{vc}
\DeclareMathOperator{\fr}{fr}
\DeclareMathOperator{\tw}{tw}

\crefname{figure}{Figure}{Figures}  
\crefname{subfigure}{Figure}{Figures}  
\crefname{observation}{Observation}{Observations}  
\crefname{theorem}{Theorem}{Theorems}
\crefname{definition}{Definition}{Definitions} 
\crefname{lemma}{Lemma}{Lemmata}
\crefname{align}{Eq}{Equalitis}
\crefname{algorithm}{Algorithm}{Algorithms}
\crefname{section}{Section}{Sections}

\begin{document}
\title{Binary $k$-Center with Missing Entries: Structure Leads to Tractability}
\author{Farehe Soheil \footnote{farehe.soheil@hpi.de}, Kirill Simonov\footnote{kirill.simonov@gmail.com}, Tobias Friedrich\footnote{tobias.friedrich@hpi.de} \\[.075cm] \textit{\small Hasso Plattner Institute, University of Potsdam, Germany}}

\maketitle
    \begin{abstract}
    $\kC$ clustering is a fundamental classification problem, where the task is to categorize the given collection of entities into $k$ clusters and come up with a representative for each cluster, so that the maximum distance between an entity and its representative is minimized.
In this work, we focus on the setting where the entities are represented by binary vectors with missing entries, which model incomplete categorical data. This version of the problem has wide applications, from predictive analytics to bioinformatics.

Our main finding is that the problem, which is notoriously hard from the classical complexity viewpoint, becomes tractable as soon as the known entries are sparse and exhibit a certain structure. Formally, we show fixed-parameter tractable algorithms for the parameters vertex cover, fracture number, and treewidth of the row-column graph, which encodes the positions of the known entries of the matrix.
Additionally, we tie the complexity of the 1-cluster variant of the problem, which is famous under the name Closest String, to the complexity of solving integer linear programs with few constraints.
This implies, in particular, that improving upon the running times of our algorithms would lead to more efficient algorithms for integer linear programming in general.
    \end{abstract}
    \section{Introduction}
    \label{sec:intro}

Clustering is a fundamental problem in computer science with a wide range of applications \citep{hansen1997cluster, hsu1979easy, shi2000normalized, ge2008joint, tan2013data}, and has been thoroughly explored \citep{1056489, baker2020coresets,kar2021feature, cohen2022towards, wu2024new, bandyapadhyay2024coresets}.
In its most general formulation, given \(n\) data points, the aim of a clustering algorithm is to partition these points into groups, called clusters, based on the similarity. 
The degree of similarity or dissimilarity between points is modeled by a given distance function.
Depending on the representation of the data points, several computationally different variants of the clustering problem arise. Commonly, the points are embedded in the $d$-dimensional Euclidean space $\mathbb{R}^d$ with the standard Euclidean distance or the distance given by $L_p$ norms, or in the space of binary strings equipped with the Hamming distance, or in a general metric space where the distance function is given explicitly. 

Additionally, there exist different approaches to how the similarity is aggregated. For the purpose of this work, we focus on the classical center-based clustering objectives.
In $k$-Center clustering, given a set of data points and the parameter $k$, the objective is to 
partition the points into $k$ clusters and identify for each cluster a point called the \emph{center}, so that the \emph{maximum} distance between the center and any point within its cluster is minimized. $k$-Median clustering is defined in the same way, except that the \emph{sum} of distances between the data points and their respective centers is minimized, and $k$-Means clustering aims to minimize the sum of \emph{squared} distances instead.

Unfortunately, virtually all versions of clustering are computationally hard, in the classical sense.
$k$-Means in $\mathbb{R}^d$ is NP-hard even on the plane (dimension $d = 2$)~\citep{MahajanNV09}, and it is also NP-hard for $k = 2$ clusters even when the vectors have binary entries~\citep{AloiseDHP09,Feige14b}.
$k$-Median and $k$-Center are NP-hard in $\mathbb{R}^2$ as well~\citep{MegiddoS84}.
Moreover, $k$-Center on binary strings under the Hamming distance is NP-hard even for $k = 1$ cluster; that is, the problem of finding the binary string that minimizes the maximum distance to a given collection of strings is already NP-hard~\citep{frances1997covering, lanctot2003distinguishing}.
The latter problem is well-studied in the literature under the name of \CS, given its importance for applications ranging from coding theory~\citep{kochman2012adversarial} to bioinformatics~\citep{Stojanovic97}.

In order to circumvent the general hardness results and simultaneously increase the modeling power of the problem, we consider the following variant of clustering with missing entries. We assume that the data points are represented by vectors, where each entry is either an element of the original domain (e.g., in $\mathbb{R}$ or $\{0, 1\}$), or the special element ``?'', which corresponds to an unknown entry.  For the clustering objective, the distance is computed normally between the known entries, but the distance to an unknown entry is always zero. 
In this way, we can define the problems \kCME and \kMME.
For formal definitions see~\cref{sec:prelim}.

These problems have a wide range of applications. For an example in predictive analytics, consider the setting of the classical Netflix Prize challenge\footnote{The problem description and the dataset is available at \url{https://www.kaggle.com/netflix-inc/netflix-prize-data}.}. The input is a collection of user-movie ratings, and the task is to predict unknown ratings. The data can be represented in the matrix form, where the rows correspond to the users and the columns to the movies, and naturally most of the entries in this matrix would be unknown. Clustering in this setting is then an important tool for grouping/labelling similar users or similar movies, based on the available data. 

Clustering with missing entries is also closely related to string problems that arise in bioinformatics applications.
In the fundamental genome phasing problem (known also as ``haplotype assembly''), the input is a collection of \emph{reads}, i.e., short subsequences of the two copies of the genome, and the task is to reconstruct both of the original sequences. Finding the best possible reconstruction in the presence of errors is then naturally modeled as an instance of \kMME with $k = 2$, where the data points correspond to the individual reads, using missing entries to mark the unknown parts of each read; the target centers represent the desired complete genomic sequences; and the clustering objective represents the total number of errors between the known reads and the desired sequences, which needs to be minimized.
In fact, \cite{Patterson2015} use exactly this formalization of the phasing problem (under the name of ``Weighted Minimum Error Correction'') as the algorithmic core of their WhatsHap phasing software.
Note that in both examples above, the known entries lie in a finite, small domain. For the technical results in this work, we focus on vectors where the vectors have binary values, e.g., in $\{0, 1\}$; the results however can be easily extended to the bounded domain setting.

Generally speaking, \kCME and \kMME cannot be easier than their fully-defined counterparts. While greatly increasing the modeling power of the problem, the introduction of missing entries poses also additional technical challenges. In particular, most of the methods developed for the standard, full-information versions of clustering  are no longer applicable, since the space formed by vectors with missing entries is not necessarily metric: the distances may violate triangle inequality. On the positive side, one can observe that in practical applications the structure of the missing entries is not completely arbitrary. In particular, the known entries are often \emph{sparse}---for example, in the above-mentioned Netflix Prize challenge, only about 1\% of the user-movie pairs have a known rating. Therefore, the ``hard'' cases coming from the standard fully-defined versions of $k$-Center/$k$-Means are quite far from the instances arising in applications of clustering with missing entries.
This motivates the aim to identify tractable cases of \kCME based on the structure of the missing entries, since the general hardness results for $k$-Center are not applicable in this setting.

Formally, we use the framework of \emph{parameterized complexity} in order to characterize such cases.
We are looking for algorithms that run in time $f(t) \cdot \poly(n)$, where $t$ is a \emph{parameter} associated with the instance, which could be any numerical property of the input, and $f(\cdot)$ is some function of this parameter.
That is, the running time may be exponential in the parameter $t$, but needs to be polynomial in the size of input for every fixed $t$.
Such algorithms are called fixed-parameter tractable (FPT), and this property heavily depends on the choice of the parameter $t$.
On the one hand, the parameter should capture the ``complexity'' of the instance, allowing for FPT algorithms to be possible; on the other hand, the parameter should be small on a reasonably broad class of instances, so that such an algorithm is applicable.
We refer to standard textbooks on parameterized complexity for a more thorough introduction to the subject~\citep{DowneyFellows13,CyganFKLMPPS15}.

\begin{figure}[h]
    \centering
    \includegraphics[width=0.5\columnwidth]{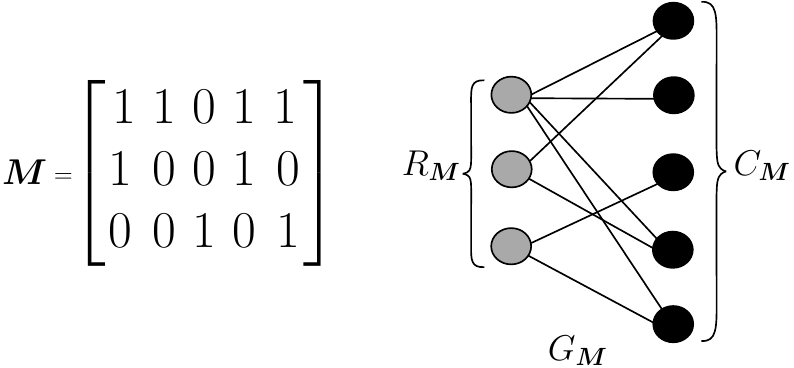}
    \caption{On the left, the mask matrix $\M$ and on the right, its corresponding incidence graph. The \emph{row} vertices, $R_{\M}$, are in gray and the column vertices $C_{\M}$ are in black.}
    \label{fig:incidence_graph}
\end{figure}

In order to apply the existing machinery and to put the parameters we consider into perspective, we encode the arrangement of the missing entries into a graph.
We say that the \emph{incidence graph} of a given instance is the following bipartite graph: the vertices are the data points and the coordinates, and the edge between a point and a coordinate is present when the respective entry is \emph{known}. When interpreting the input as a matrix, where the data points are the rows, replacing the known entries by ``1'' and missing entries by ``0'' results exactly in the biadjacency matrix of the incidence graph. We call this matrix the \emph{mask} matrix of the instance, denoted by $\M$, and denote the incidence graph of the instance by $G_{\M}$; see \Cref{fig:incidence_graph} for an illustration.

We mainly consider the following three fundamental sparsity parameters of the incidence graph $G_{\M}$:
\begin{itemize}
    \item vertex cover number $\vc(G_{\M})$, which is the smallest number of vertices that are necessary to cover all edges of the graph;
    \item fracture number $\fr(G_{\M})$, which is the smallest number of vertices one needs to remove so that the connected components of the remaining graph are small, i.e., their size is bounded by the same number;
    \item treewidth $\tw(G_{\M})$, which is a classical decomposition parameter measuring how ``tree-like'' the graph is.
\end{itemize}  
See \Cref{fig:vertex_cover_graph,fig:fracture_matrix} for a visual representation of instances with small vertex cover and fracture number, respectively. Treewidth is the most general parameter out of the three, and it encompasses a wide range of instances.
For example, the main algorithmic ingredient in the WhatsHap genomic phasic software~\citep{Patterson2015} is the FPT algorithm for \kMME parameterized by the maximum number of known entries per column\footnote{Called \emph{coverage} in their work.}, in the special case where $k = 2$ and the known entries in each row form a continuous subinterval; treewidth of the incidence graph is never larger than this parameter.
Vertex cover, on the other hand, is the most restrictive of the three; however, comparatively small vertex cover may be a feasible model in settings such as the Netflix Prize challenge, where most users would only have ratings for a relatively small collection of the most popular movies.
The fracture number aims to generalize the setting of small vertex cover, to also allow for an arbitrary number of small local ``information patches'', outside of the few ``most popular'' rows and columns. Note that fracture number is a strictly more general parameter than vertex cover number, since removing any vertex cover from the graph results in connected components of size one; that is, for any instance, $\fr(G_{\M}) \le \vc(G_{\M})$. It also holds that $\tw(G_{\M}) \le 2\fr(G_{\M})$, see \Cref{sec:fr_fpt} for the details.


Previously, the perspective outlined above has been successfully applied for \kMME. \citep{GanianHKOS22} show that the problem admits an FPT algorithm parameterized by treewidth of the incidence graph in case of the bounded domain, as well as further FPT algorithms for real-valued vectors in more restrictive parametrization.
However, similar questions for \kCME remain widely open.

In this work, we aim to close this gap and investigate parameterized algorithms for the $k$-Center objective on classes of instances where the known entries are ``sparse'', in the sense of the structural parameters above.
Our motivation stems from the following.
First, $k$-Center is a well-studied and widely applicable similarity objective, which in certain cases might be preferable over $k$-Means; specifically, whenever the cost of clustering is associated with each individual data point and has to be equally small, as opposed to minimizing the total, ``social'', cost spread out over all data points.
Second, $k$-Center is interesting from a theoretical perspective, being a natural optimization target that on the technical level behaves very differently from $k$-Means. Our findings, as described next, show that the $k$-Center objective is in fact more challenging than $k$-Means in this context, and we prove that
\kCME is as general as a wide class of integer linear programs.

%

\subparagraph*{\textbf{Our contribution.}}
We present several novel parameterized algorithms for \kCME. First, we show that \kCME is FPT when parameterized by the vertex cover number $\vc(G_{\M})$ of the incidence graph plus $k$. 
Specifically, we prove the following result. Here and next, $n$ is the number of data points in the instance and $m$ is their dimension.

\begin{restatable}{theorem}{thmvc}
    \label{thm:vc}
    \kCME admits an algorithm with running time $$2^{\bo{k \cdot \vc(G_{\M}) + \vc(G_{\M})^2 \cdot\log{\vc(G_{\M})}}} \poly(nm).$$
\end{restatable}
This result can be compared to the result of \citep{EibenGKOS23}, who considered the complementary parametrization of the same problem (under the name of \textsc{Any-Clustering-Completion}). That is, they consider the minimum number of rows and columns that are needed to cover all \emph{missing} entries.
Interestingly, for their FPT algorithm, it was also necessary to include the target distance $d$ in the parameter---we do not need this restriction in our setting, which highlights the property that instances of \kCME, where missing entries are dense, are, in a sense, easier.

Moving further, we extend the result of \Cref{thm:vc} to the more general setting where the parameter is the fracture number of the incidence graph $G_{\M}$.
We show that one can achieve the running time of \Cref{thm:vc} even for fracture number, which is the most technical result of this paper.

\begin{restatable}{theorem}{thmfr}
 \label{thm:fr}
    \kCME admits an algorithm with running time $$2^{\bo{k \cdot \fr(G_{\M}) + \fr(G_{\M})^2\cdot \log{\fr(G_{\M})}}} \poly(nm).$$
\end{restatable}

To the best of our knowledge, no previous work on clustering problems considers the fracture number as the parameter; however it has been successfully applied to other fundamental problems such as Integer Linear Programs (ILPs)~\citep{GavenciakKK22} and Edge Disjoint Paths~\citep{GanianOR21}. Furthermore, a very similar parameter, equivalent to fracture number, has been studied in the literature under the name \emph{vertex integrity}, for example in the context of Subgraph Isomorphism~\citep{BodlaenderHKKOO20} and algorithmic metatheorems~\citep{LampisM24}. 

In order to prove \Cref{thm:fr}, we also need an algorithm with respect to the treewidth of the incidence graph $G_{\M}$, stated in the next theorem. We denote by $d$ the target radius of the cluster, i.e., the maximum distance between a point and its cluster center.

\begin{restatable}{theorem}{thmtw}
    \label{thm:tw}
    \kCME admits an algorithm with running time $$d^{\bo{\tw(G_{\M})}} 2^{\bo{k \cdot \tw(G_{\M})}} \poly(nm).$$
\end{restatable}

In other words, the problem is FPT when parameterized by $\tw(G_m) + d + k$, or XP when parameterized by $\tw(G_m) + k$. Note that, as opposed to \Cref{thm:vc} and \Cref{thm:fr}, here we need the dependence on $d$ in the exponential part of the running time. This is, however, still sufficient to enable the algorithm claimed by \Cref{thm:fr}.

While we are not aware of matching hardness results based on standard complexity assumptions, we can nevertheless argue that improving the running time in \Cref{thm:vc,thm:fr} resolves a fundamental open question.
Specifically, we show a parameterized equivalence between \CS, parameterized by the number of strings, and Integer Linear Program (ILP) with bounded variables, parameterized by the number of rows.
While we use the reduction from \CS to ILP as a building block in the algorithm of \Cref{thm:vc}, the reduction in the other direction, i.e., from ILP to \CS, is most relevant here. Formally, it yields the following statement:
\begin{restatable}{theorem}{thmilp}
       For any $\alpha > 0$, assume that \CS admits an algorithm with running time $2^{\bo{n^{1 + \alpha}}} \cdot \poly(n\ell)$, where $n$ is the number of strings and $\ell$ is their length. Then the ILP $\{\A \x = \bm{b}: \forall i, \,\ell_{i} \le x_i \le u_i \}$, where $\A \in \mathbb{Z}^{r \times c}$, $\mathbf{b} \in \mathbb{Z}^r$, and $\ell_i \leq u_i \in \Z$, can be solved in time $2^{\bo{r^{1 + \alpha + o(1)}}} \cdot \poly(rc)$, assuming that $\norm{\A}{\infty} = \bo{r}$ and $\delta = 2^{\bo{r}}$, where $\delta=\max_{i \in [c]} (u_i - \ell_{i})$. 
        \label{thm:ilp}
\end{restatable}
Note that \CS is a special case of \kCME with $k = 1$ and no missing entries, where also the number of strings matches with the vertex cover of the known entries. Therefore, we immediately get that improving the $\vc(G_{\M})^2$ term in the exponent of 
\Cref{thm:vc} to $\vc(G_{\M})^{1 + \alpha}$ implies the same improvement for this class of ILP instances, and the same holds for the result of \Cref{thm:fr}. Notably, the result of \Cref{thm:ilp} was independently discovered by \citep{RohwedderW2024} in a recent preprint; they also provide further examples of problems, where such an improvement implies breaking long-standing barriers in terms of the best-known running time, and conjecture that this might be impossible.

\subparagraph*{\textbf{Related work.}}
Clustering problems are also extensively studied from the approximation perspective.
The $k$-Center problem classically admits 2-approximation in poly-time~\citep{gonzalez1985clustering, feder1988optimal}.
On the other hand, approximating $\kC$ in $\mathbb{R}^2$ within a factor of 1.82 is NP-hard~\citep{mentzer2016approximability, chen2021mentzer}.

\CS is well-studied under various parameters~\citep{LiMW02,gramm2003fixed,MaS08,ChenMW14} and in the area of approximation algorithms~\citep{gkasieniec1999efficient, li2002closest, ma2010more, mazumdar2013chebyshev}. Abboud et al.~\citep{abboud2023can} have shown that $\CS$ on binary strings of length $\ell$ can not be solved in time $(2 - \varepsilon)^{\ell} \cdot \poly(n \ell)$ under the SETH, for any $\varepsilon > 0$. This can be seen as a tighter analogue of \Cref{thm:ilp} for the parametrization of \CS by the lengths of the strings. 
A version of the \CS with wildcards was studied from the viewpoint of parameterized complexity~\citep{HermelinR15}; the wildcards behave exactly like missing entries in our definition, therefore this problem is equivalent to \kCME for $k = 1$.

\citep{KnopKM20} studied combinatorial $n$-fold integer programming, yielding in particular $n^{\bo{n^2}}\poly(m)$ algorithms for \CS, \CS with wildcards, and \textsc{$k$-Center} (with binary entries), where $n$ is the number of strings/rows and $m$ is the length of the strings/number of coordinates. These results can be seen as special cases of our Theorem~\ref{thm:fr}.

The versions of \kCME/\kMME with zero target cost of clustering is considered in the literature under the class of ``matrix completion'' problems. There, given a matrix with missing entries, the task is to complete it to achieve certain structure, such as few distinct rows (resulting in the same objective as in the clustering problems) or small rank. Parameterized algorithms for matrix completion problems were also studied~\citep{GanianKOS18}.

Finally, \kMME has been studied in $\mathbb{R}^d$ from a parameterized approximation perspective. Geometrically, each data point with missing entries can be seen as an axis-parallel linear subspace of $\mathbb{R}^d$, and the task is to identify $k$ centers as points in $\mathbb{R}^d$ that minimize the total squared distance to the assigned subspaces. \citep{EibenFGLPS21} show that this problem admits a $(1 + \varepsilon)$ approximation in time $2^{\poly(k, \varepsilon, \Delta)} \cdot \poly(nd)$, where $\Delta$ is the maximum number of missing entries per row, meaning that the subspaces have dimension at most $\Delta$.

\subparagraph*{\textbf{Paper organization.}} We define the necessary preliminaries in \Cref{sec:prelim}. Then we show the reduction from \CS to ILP in \Cref{sec:CS_to_ILP}, and the reduction from ILP to \CS in \Cref{sec:ILP_to_CS}. \Cref{sec:vc_fpt,sec:tr_fpt,sec:fr_fpt} are dedicated to the respective FPT results for the parameters vertex cover, treewidth and fracture number. We conclude in \Cref{sec:conclusion}.
Due to space constraints, most of the technical proofs are deferred to the appendix.
    \section{Preliminaries}
     \label{sec:prelim}
    In this section, we introduce key definitions and notations used throughout the paper.
For an integer $n$ we write $[n]$ to denote the set $\{1, \dotso, n\}$. We use $\Z_{+}$ to denote the set of non-negative integers.
For a vector of real numbers $\bm{v} \in \R^{\ell}$ with length $\ell$,  its $i$-th entry is denoted by $\bm{v}[i]$ and $v_i$ interchangeably. 
Similarly, for a binary string $s \in \{0,1\}^{\ell}$ of length $\ell$, its $i$-th bit is denoted by $s[i]$, and $s_i$ and we write $s=[s_1, \dotso, s_{\ell}]$ to denote the whole string. 
For a set $I \subseteq [\ell]$, the substring of $s$ restricted to the indices in $I$ is written as $s[I] \in \{0, 1\}^{|I|}$.
For two indices $i,j \in [\ell]$, the substring from $s[i]$ throughout $s[j]$ (inclusive) is denoted by $s[i, j] \in \{0, 1\}^{j-i+1}$.
The complement of $s$ is represented by $\overline{s}$.
The Hamming distance between two binary strings $s_1, s_2 \in \{0,1\}^{\ell}$ is denoted by $\hd: \{0,1\}^{\ell} \times \{0,1\}^{\ell} \rightarrow \Z_{+} $ with $\hd(s_1, s_2) = \big| \big\{i \in [\ell] : s_1[i] \neq s_2[i] \big\} \big|$. 
For a graph $G$, the set of vertices and edges are denoted by $V(G)$ and $E(G)$ respectively.\\
Consider a matrix \( \A \) with \( n \) rows and \( m \) coordinates.
The sets of rows and coordinates of \( \A \) are denoted by \( R_{\A} = \{r_1, r_2, \dotso, r_n\} \) and \( C_{\A} = \{c_1, c_2, \dotso, c_m\} \) respectively.
For $i \in [n]$ and $j \in [m]$, the $i$-th row of $\A$ is denoted by $\A[i]$ and the $j$-th entry of $\A[i]$ is written as $\A[i][j]$.
We may also use $a_i$ and $a_{ij}$ respectively to refer to $\A[i]$ and $\A[i][j]$, when clear from context.
For matrices \( \A \) and \( \B \), their entry-wise subtraction is written as \( \A - \B \), meaning \( (\A - \B)[i][j] = \A[i][j] - \B[i][j] \). 
Similarly, their entry-wise product is denoted by \( \A \circ \B \), where \( (\A \circ \B)[i][j] = \A[i][j] \cdot \B[i][j] \).
For two binary matrices $\A$ and $\B$, we define their row-wise Hamming distance as a vector $\hd\big(\A, \B\big)$, i.e, $\hd \big(\A,\B\big) \big[ i \big] = \hd\big(\A[i], \B[i] \big)$.
Moreover, a missing entry is denoted by ``?''.

For a binary matrix $\M \in \{0,1\}^{n \times m}$, its \emph{incidence graph} is an undirected bipartite graph defined as $G_{\M}=(V_{\M}, E_{\M})$, where $V_{\M}=(R_{\M} \cup C_{\M})$ and $E_{\M} = \big \{(u,v): u \in R_{\M}, \,v \in C_{\M}, \, \M[u][v]=1 \big \}$.
We use \emph{row }(resp. \emph{coordinate}) vertices to represent the vertices in $G_{\M}$ that correspond to the rows (coordinates) of $\M$.
For a visual representation of the incidence graph, refer to Figure~\ref{fig:incidence_graph}. 

With these definitions, we can now proceed to formalize the main problem in question.
\begin{definition}[\kCME] Given matrices $\A \in \{0, 1, ?\}^{n \times m}$, and $\M \in \{0,1\}^{n \times m}$, an integer $k$ and a column vector $\bm{d}=(d,d,\dotso,d) \in \{d\}^n$. The task is to determine if there exists a binary matrix $\B \in \{0,1\}^{n \times m}$ with at most $k$ distinct rows such that
\[
\hd\big((\M \: \circ \: \A), (\M \: \circ \:\B)\big) \leq \bm{d}.
\]
\end{definition}
Note that in the context of a \kCME instance, we use the term ``point'' interchangeably with ``row'' and ``coordinate'' interchangeably with ``column'', since the points to be clustered are represented as rows of the matrices $\A$ and $\M$.

We also define here the related problems.
\begin{definition}[\CS] Given a set of $n \in \Z_{+}$ binary strings $S=\set{s_1}{s_n}$ each of length $\ell$ and a non-negative integer $d \in \Z_{+}$, determine whether there exist a binary string $s$ of length $\ell$ such that for all $i \in [n]$, $$\hd(s, s_i) \leq d.$$
\end{definition}

\begin{definition}[\NUCS] Given a set of $n \in \Z_{+}$ binary strings $S=\set{s_1}{s_n}$ each of length $\ell$ and a distance vector $\bm{d} = \big(d_1, \dotso, d_n \big) \in \Z_{+}^n$, determine whether there exist a binary string $s$ of length $\ell$ such that for all $i \in [n]$, $$\hd(s, s_i) \leq d_i.$$
\end{definition}

\begin{definition}[\ILP] Given a constraint matrix $\A \in \mathbb{Z}^{r \times c}$, a column vector $\mathbf{b} \in \mathbb{Z}^r$, and integers $\ell_i \le u_i$ for each $i \in [c]$, 
determine whether there exists a vector $\x \in \mathbb{Z}^c$ such that
$$\A \cdot \x = \bm{b}$$ and $\ell_i \le x_i \le u_i$ for each $i \in [c]$.

\end{definition}
     \section{Reduction from $\CS$ to $\ILP$}
     \label{sec:CS_to_ILP}
In this section, we show an ILP formulation of the \CS problem, which will be used in later sections as a building block for our algorithms. Given an instance $I=(S,d)$ of the $\CS$ problem where $n \in \Z_{+}$ denotes the number of binary strings in $S=\set{s_1}{s_n}$, each of length $\ell$, the objective is to determine whether there exists a binary string $s$ of length $\ell$, such that $\max_{i} \big( \hd (s, s_i)\big) \leq d$.
We formulate this problem as an $\ILP$ instance with $p$ variables and $n$ constraints where $p = \min(\ell, 2^n)$.
\begin{restatable}{theorem}{thmcstoilp}
An instance $I=(S,d)$ of the $\CS$ problem with $n$ binary strings of length $\ell$, can be reduced to an $\ILP$ instance $\big\{\A \cdot \x \leq \bm{b}: \forall i,\, 0 \leq x_i \leq u_i \big\}$ where $\A \in \{-1, 1\}^{n \times p}$ and $\bm{b} \in [d]^n$ with $p \leq \min (\ell, 2^n)$, in time $\bo{n \cdot \ell}$. 
     \label{thm:CS_to_ILP}
\end{restatable}
\begin{proof}
To begin with, observe that the input strings can be represented as a binary matrix $\M$.
This matrix has $n$ rows and $\ell$ coordinates, where for $i\in [n]$, $\M[i]$ corresponds to the string $s_i$, and for $ j\in [\ell] $,  the entry $\M[i][j]$ represents the $j$-th character of the string $s_i$. 
For each coordinate \(j \in [\ell]\), let \(c_j\) denote the binary string of length $n$ induced by the \(j\)-th coordinate across all strings in $\M$, that is $c_j = \M[-][j]$.  
We call $c_j$, the $j$-th column of $\M$.
Note that, each $c_j$ can take one of up to \(2^n\) distinct binary string configurations.
However, certain configurations may repeat across different $c_j$'s, meaning multiple instances of $c_j$ may share the same binary string value.
Additionally, some potential configurations might not be realized by any $c_j$.
Thus, while there are $2^n$  possible binary strings, not all will necessarily appear, nor will they be unique across all $c_j$'s.
To illustrate this more clearly, consider the input set of binary strings $S=\{\text{\say{0110110}, \say{1001001}, \say{1011011}, \say{1111111}}\}$.
The corresponding binary matrix $\M$ is depicted in \cref{fig:reordered_column_matrix} (a).
The columns $c_1$, $c_4$, and $c_7$ are of type $\qt{0111}$ and $c_2 = \qt{1001}$ provided as examples of the strings induced by coordinates 1, 4 and 7 of $\M$, respectively.
\begin{figure}[htbp]
    \centering
    \includegraphics[width=0.5\columnwidth]{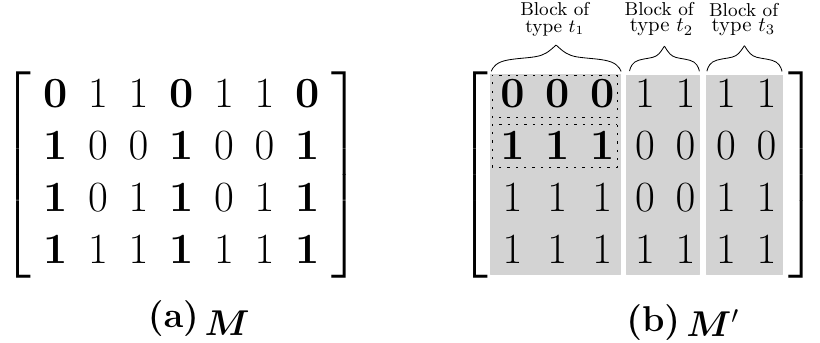}
    \caption{On the left is matrix $\M$, with each row corresponding to a string in $S$. Columns $c_1$, $c_4$, and $c_7$ are represented in bold. The \textit{reordered} input matrix $\M'$ in on the right. Strings of the same \textit{type}, are grouped as \textit{blocks}, with each block colored in gray. In each block, the rows are either consecutive ones or consecutive zeros. The first and second row restricted to the block of type $t_1$, represented in bold font, are \text{``000''}and \text{``111''} respectively}
    \label{fig:reordered_column_matrix}
\end{figure}
Let $T=\set{t_1}{t_{p}} = \{c_j |\, j \in [\ell]\}$, denote the set of all binary strings of length $n$ that appear as the columns of $\M$, ordered in increasing value based on their binary representation.
Note that $p \leq 2^n$.
We refer to each $t_i$ as a \emph{column type}. 
For each column type $t_i$, define $n_{i}$ as the number of coordinates $j$ in $\M$, where the column $c_j$ is of type $t_i$.
In other words, $n_{i}=  |\big\{j: \, j \in [\ell] , c_j = t_i  \big\}|$ with $\sum_{i=1}^{p} n_{i} = \ell$. 
For each $i \in [p]$, the \emph{range} of column type $t_i$ is defined as $[\ell_i, r_i]$ where
$$\ell_i = \left( \sum_{j=1}^{i-1} n_j \right) + 1 \quad \text{and} \quad r_i = \sum_{j=1}^{i} n_j.$$
In Figure\ref{fig:reordered_column_matrix}(a), column types are $T = \{ \qt{0111},  \qt{1001},  \qt{1010}\}$ with $n_{1}=3$, $n_{2}=2$, and $n_{3}=2$. 
Additionally, the ranges of the column types in Figure\ref{fig:reordered_column_matrix}(a) are given by $[\ell_1,\, r_1]=[1, \, 3]$, $[\ell_2,\, r_2]=[4, \, 5]$, and $[\ell_3,\, r_3]=[6, \, 7]$.
Intuitively, the range $[\ell_i, r_i]$ of a column type $t_i$, specifies the interval where columns of type $t_i$ are placed after the \textit{reordering} of matrix $\M$, which will be detailed in the following explanation.\\
Define the permutation \(\pi: [\ell] \rightarrow [\ell]\) as a rearrangement of the coordinates of the matrix $\M$, such that all columns $c_j$, for $j \in [\ell]$, that share the same column type $t_i$ (i.e. $c_j = t_i$) are positioned consecutively between the coordinates $\ell_i$ and $r_i$.
Let $\M'$ denote the \textit{reordered} matrix obtained by applying the permutation $\pi$ to the coordinates of matrix $\M$.
Then, \textit{blocks} are defined as contiguous groups of columns in $\M'$ that correspond to the same type. 
Thus there are $p$ blocks in matrix $\M'$.
These blocks of columns appear in sequence according to the binary value of their types $\set{t_1}{t_{p}}$, with the size of each block determined by the corresponding value $n_i$.
More formally, for each type $t_i$ with $i \in [p]$ and for every $j \in [\ell_i, r_i]$, it holds that $\M'[-][    j] = t_i$.
In other words, \(\pi\) rearranges the coordinates of \(\M\) so that, each column $c'_j$ of $\M'$ with $j \in [\ell_i, r_i]$, induces a string of type \(t_i\).
Applying $\pi$ to coordinates of matrix $\M$ in Figure \ref{fig:reordered_column_matrix} (a), results in matrix $\M'$ depicted in Figure \ref{fig:reordered_column_matrix} (b).
For convenience, we write $\pi(S)$ to denote the application of permutation $\pi$ to a set of strings $S$, and $\pi(s_i)$ when applying it to a single string $s_i$.
With the necessary tools and the permutation $\pi$ outlined above, it is easy to verify the following observation:
\begin{observation}
    Assume $s \in \{0,1\}^{\ell}$ is a solution to the given instance $I=(S,d)$ of binary $\CS$, then $\pi(s)$ is a solution to the permuted instance $I'=(S',d)$ where $S'=\pi(S)$. \label{obs:reordered_matrix}
\end{observation}
As a direct result of \cref{obs:reordered_matrix}, we can now address the permuted instance $I'=(S',d)$, where $S'=\set{s'_1}{s'_n}$, rather than the original instance $I=(S,d)$.
Let $s^*$ be a solution to $I'$.
For each input string $s'_k \in S'$, we can express the Hamming distance between $s'_k$ and $s^*$ as follows:
\begin{align}
    \hd \big(s^*, s'_k\big) = \sum_{i=1}^{p} \hd \big(s^*[\ell_i, r_i] , s'_k[\ell_i, r_i]\big) 
\label{eq:hd_all}
\end{align}
Let $\M'$ denote the permuted matrix consisting of the strings in $S'$.
It is important to note that, by the definition of the permutation $\pi$, for any $i\in [p]$,  column type $t_i$ appears consecutively $(r_i-\ell_i+1)$ times  throughout the coordinates $\left[ \ell_i, r_i \right]$.
As a result, the rows restricted to every block $i \in [p]$, are either strings of consecutive zeros or consecutive ones, each of length $(r_i-\ell_i+1)$.
In simpler terms, the rows of $\M'$ restricted to the coordinates in $\left[ \ell_i, r_i \right]$ (i.e the range associated with column type $t_i$), consist of either $(r_i-\ell_i+1)$ consecutive zeros or $(r_i-\ell_i+1)$ consecutive ones.
For further illustration, refer to \cref{fig:reordered_column_matrix}.
Thus, for every input string $s'_k \in S'$, the Hamming distance between $s^*$ and $s'_k$ along the coordinates in $\left[\ell_i, r_i \right]$ is determined by the number of zeros and ones in $s^*\left[\ell_i, r_i \right]$. 
Therefore, we can disregard specific positions of ones and zeros within $s^*\left[\ell_i, r_i \right]$, and only take into account the number of their occurrence.\\
For each column type $t_i$, let $z_i = |\big\{j \in [\ell_i, r_i]: s^*[j] = 0\big\}|$ denote the number of zeros that will appear in $s^*\left[ \ell_i, r_i \right]$.
Then, for each $i \in [p]$ and $s'_k \in S'$, it follows that:
\begin{align}
        \hd \big(s^*[\ell_i, r_i] , s'_k[\ell_i, r_i]\big) = 
        \begin{cases}
              z_i & \text{if $s'_k[\ell_i,r_i]=\Bar{1}$}\\
              n_i-z_i & \text{if $s'_k[\ell_i,r_i]=\Bar{0}$}.
        \end{cases} 
        \label{eq:hd_with_ variables}
\end{align}
Define $y_i:S'\rightarrow \{0,1\}$, for $i \in [p]$, as follows:
\begin{align}
            y_i(s'_k) = 
        \begin{cases}
              0 & \text{if $s'_k[\ell_i,r_i]=\Bar{1}$}\\
              1 & \text{if $s'_k[\ell_i,r_i]=\Bar{0}$}.
        \end{cases} 
        \label{eq:aux_mapping}
\end{align}
Let $\mu_k = \sum_{i=1}^{p} y_i(s'_k) \cdot n_i$ denote the number of zeros in $s'_k$.
Then using \cref{eq:hd_all}, \cref{eq:hd_with_ variables}, and \cref{eq:aux_mapping}, the Hamming distance between $s^*$ and $s'_k \in S'$ can be expressed as follows:
\begin{align}
    \hd \big(s^*, s'_k\big) &= \sum_{i=1}^{p} y_i(s'_k) \cdot n_i + (-1)^{y_i(s'_k)} \cdot z_i \notag \\
    &= \mu_k + \sum_{i=1}^{p}(-1)^{y_i(s'_k)}\cdot z_i
\end{align}
To put it simply, consider now an $\ILP$ instance where $z_i$'s, for $i \in [p]$, form the variables subject to the bounds $0 \leq z_i \leq n_i $.
Moreover, for each input string $s'_k$, there is a corresponding constraint of the form $d_H(s^*, s'_k)\leq d$ based on the equivalence in \cref{eq:aux_mapping}.
For every $s'_k \in S'$, this results in a constraint of form $\mu_k + \sum_{i=1}^{p}(-1)^{y_i(s'_k)}\cdot z_i \leq d$.
However, these constraints are not yet in the standard form. 
The reason is that an integer $\mu_k$ appears on the left-hand side of each constraint.
The next step, therefore, is to shift this integer to the right-hand side while retaining the variables on the left.
This transformation, yields an $\ILP$ instance represented as:
\begin{align}
    \big\{\A\cdot\bm{z}\leq \bm{b} : \, \bm{z} \in \Z^{p}, \, 0 \leq z_i \leq n_i \big\}
    \label{eq:CS_ILP_formulation}
\end{align}
where, for $i \in [n]$ and $j \in [p]$, the constraint matrix $\A \in \{-1, 1\}^{n \times p}$ is defined by $\A[i][j] = (-1)^{y_j(s'_i)}$, and the constant matrix $\bm{b} \in \Z^n$, is given by $\bm{b}[i] = d - \mu_i$.
Note that all the columns types, their ranges and sizes can be determined in time $\bo{n \cdot \ell}$.
Also, the values of $\mu_i$, for $i \in [p]$ and $y_j(s_i)$, for $j \in [p]$ can be computed with the same running time.
\end{proof}
In \citep{eisenbrand2019proximity}, Eisenbrand and Weismantel designed a fast dynamic programming approach that solves an $\text{ILP}$ instance of the form 
\begin{equation}
    \label{eq:upperbounded_ilp}
    \max\{\bm{c}^T \, \x:  \A \x = \bm{b},\, 0 \leq \x \leq \bm{u}, \, \x \in \Z^n \}
\end{equation}
where $\A \in \Z^{m\times n}$ with $\norm{\A}{\infty} \leq \Delta$, $\bm{b} \in \Z^m$, $\bm{u} \in \Z_{+}^n$, and $\bm{c} \in \Z^{n}$ in time
\begin{equation}
n\cdot \bo{m}^{{(m+1)}^2} \cdot \bo{\Delta}^{m \cdot (m+1)} \cdot \log^2{m \cdot \Delta}.
\label{eq:ILP_running_time}
\end{equation}
Note that, by introducing a slack variable for each constraint, an $\text{ILP}$ instance of the form $\{\x: \A x \leq \bm{b},\, 0 \leq \x \leq \bm{u},\, \x\in \Z^{p} \}$ can be transferred into form \eqref{eq:upperbounded_ilp}.
Therefore, putting \cref{thm:CS_to_ILP} and \cref{eq:ILP_running_time} together, we arrive at the following corollary.
\begin{corollary}\label{cor:cs}
    An instance of \CS $I=(S, d)$ with $n$ binary strings of length $\ell$ can be solved in time $\bo{\ell}\cdot n^{\bo{n^2}}\log^2{n}$. This also holds for non-uniform distances and in the presence of missing entries.\label{co:binary_closest_string_time} 
\end{corollary}

In Corollary~\ref{cor:cs}, different distances and missing entries are accommodated by a straightforward change in the ILP formulation. Notably, running time similar to Corollary~\ref{cor:cs} has been previously shown~\citep{KnopKM20} for all listed \CS versions. The goal of this section is to state the ILP formulation explicitly.
    \section{Reduction from $\ILP$ to $\CS$}
    \label{sec:ILP_to_CS}
    This section is dedicated to the reduction from \ILP to \CS that preserves the number of rows up to a factor of $\bo{\log{\norm{\A}{\infty}}}$, where $\A$ is the constraint coefficient matrix in the \ILP instance. 
More precisely, in this section, we outline the sequence of reductions that leads to the proof of the following theorem, which is then used to establish Theorem~\ref{thm:ilp}.
\begin{restatable}{theorem}{thmhardness}
 Let $\{ \A   \x= \bm{b}:\, \forall i, \,  \ell_{i} \le x_i \le u_i\}$ be an \ILP instance, with $\A \in \Z^{r \times c}$, $\bm{b}\in \Z ^ r$, and $\ell_i, u_i \in \Z$ for each $i \in [c]$.
    In polynomial time
    , one can construct an equivalent instance of \CS with $n = \bo{\log{\norm{\A}{\infty}} \cdot r}$,
    $\ell = \bo{\norm{\A}{\infty} \cdot \log^2{\norm{\A}{\infty}}\cdot r \cdot c \cdot \delta}$,
    and $d=\bo{\norm{\A}{\infty} \cdot \log{\norm{\A}{\infty}} \cdot c \cdot \delta}$;
    here, $\delta = \max_{i \in [c]} (u_i - \ell_{i})$.
    \label{thm:hardness}
\end{restatable}
We begin by showing that a general \ILP instance, where the coefficients may be negative, and variables may have negative lower bounds, can be transformed, in linear time with respect to the number of variables, into an equivalent instance with non-negative variable domains, i.e., of form $0 \le y_i \leq u'_i$, non-negative coefficient matrix and non-negative right-hand side, while mostly preserving the number of constraints, variables and the magnitude of the entries.

\begin{restatable}{lemma}{smallerdomain}
    Let $\{ \A   \x= \bm{b}:\, \forall i,  \ell_{i} \le x_i \le u_i\}$ be an \ILP instance with $\A \in \Z^{r \times c}$, and $\bm{b}\in \Z ^ r$, an integer column vector.
    In time $\bo{r \cdot (r + c)}$, we can construct an equivalent instance of \ILP  $\{ \A' \bm{y}= \bm{b'}:\, \forall i,  0 \le y_i \le u'_i\}$, with $\bm{b'}\in \Z_{+} ^ {2r}$, $\A' \in \Z_{+}^{2r \times (r + c)}$ such that $\norm{\A'}{\infty}=\norm{\A}{\infty}$, $\norm{\bm{b'}}{\infty}=\bo{c \cdot \delta \cdot \norm{\bm{A}}{\infty}}$ and $\max \{u_i'\} = \bo{c \cdot \delta \cdot \norm{\bm{A}}{\infty}}$, where $\delta = \max_{i \in [c]} (u_i - \ell_i)$.
    \label{lemma:smallerdomain}
\end{restatable}
\begin{proof}
    To show the lemma, consider the following two transformations.    
    First, for each variable $\ell_{i} \le x_i \le u_i$, replace it with a new variable $y_i=x_i-\ell_i$. This will shift the lower bound to $0$, so the new variable $y_i$ equivalently satisfies $0 \le y_i \leq u'_i$, where $u'_i=u_i-\ell_i$.
    Note that the right-hand side in the corresponding constraints has to be changed accordingly.
    That is, each constraint $\A[j]^T \cdot \x = \bm{b}[j]$, equivalently, $\sum_{i = 1}^c \A[i][j] \cdot x_i = \bm{b}[j]$, is replaced by an equivalent constraint $\sum_{i = 1}^c \A[i][j] \cdot y_i = \bm{b}[j] - \sum_{i = 1}^c \A[i][j] \cdot \ell_i$.
    Therefore, by setting $\bm{b}''[j] = \bm{b}[j] - \sum_{i = 1}^c \A[i][j] \cdot \ell_i$,
    the \ILP instance $\{ \A \cdot \bm{y}= \bm{b}'':\, \forall i,  0 \le y_i \le u'_i\}$ is equivalent to the original instance, while having all variable domains non-negative and the same coefficient matrix $\A$.
    Also note that for each $j \in [r]$, $\bigl|\bm{b}''[j]\bigr|$ should be at most $c \cdot \delta \cdot \norm{\bm{A}}{\infty}$, as otherwise the $j$-th constraint is clearly infeasible.
    Therefore, if for some $j \in [r]$, the value of $\bigl|\bm{b}''[j]\bigr|$ exceeds the value above, return a trivial no-instance, and for the rest of the proof assume that the bound holds.

    The next step is to deal with the negative coefficients in the constraint matrix $\A$.
    To do so, consider each constraint $\A[j]^T \cdot y = \bm{b'}[j]$.
    Let $\Sigma_{-} + \Sigma_{+} = \bm{b'}[j]$, be the equivalent expression of the constraint, where $\Sigma_{-}$ collects the terms in $\A[j]^T \cdot y$ with negative coefficients and $\Sigma_{+}$ collects the terms with positive coefficients.
    Now replace the constraint $\Sigma_{-} + \Sigma_{+}=\bm{b'}[j]$ with two new constraints: $\Sigma_{+}+y_{c + i}=\bm{b}[i]+N$ and $y_{c + i} -\Sigma_{-}=N$ where $N$ is a sufficiently large positive integer: $N = c \cdot \delta \cdot \norm{\A}{\infty}$ and $\bm{y}[c + i]$ is a new variable with bounds $0 \leq y_{c + i} \leq N$.
    It is easy to see that by subtracting the second constraint from the first one, we obtain the original constraint. On the other hand, any variable assignment in the original instance fixes exactly one possible assignment for $y_{c + i}$ in each of the two new constraints.
    Therefore, these two constraints are equivalent to the original constraint. 
    Now note that in the constraint $\Sigma_{+}+y_{c + i}=\bm{b}[i]+N$, all the variables on the left-hand side have positive coefficients.
    Also, since $\Sigma_{-}$ consists only of negative terms, the left-hand side of the constraint $y_{c + i} -\Sigma_{-}=N$ has only non-negative coefficients as well.
    Moreover, if there is a negative entry in the right-hand side of any constraint after the transformation, the original \ILP instance is infeasible, so it suffices to return a trivial no-instance.
    Otherwise, the right-hand sides of all new constraints are non-negative.
    Denote by $\A'$ the coefficient matrix of the new constraints, and by $\bm{b}'$ the vector of the right-hand sides; by construction, $\A' \in \Z_{+}^{2r \times (r + c)}$, $\bm{b'}\in \Z_{+} ^ {2r}$. From the above, the \ILP instance $\{ \A' \bm{y}= \bm{b'}:\, \forall i,  0 \le y_i \le u'_i\}$ is equivalent to the original one, where $u'_i = N$ for $c + 1 \le i \le r + c$. It holds that $\norm{\A'}{\infty}=\norm{\A}{\infty}$, and by the choice of $N$ and $\bm{b}''$, $\norm{\bm{b'}}{\infty}=\bo{c \cdot \delta \cdot \norm{\bm{A}}{\infty}}$.
\end{proof}

By lemma~\cref{lemma:smallerdomain}, we observe that a general \ILP instance is equivalent to an instance where everything is non-negative, with a bounded blow-up in the number of constraints, variables and the magnitude of values. Thus, we continue our chain of reductions from a non-negative ILP.
In the next step, we reduce from arbitrary bounds on the variables to binary variables, as stated in the following lemma.

\begin{restatable}{lemma}{lemmarbtobin}
     Let $\{ \A   \x=\bm{b}: \forall i, \, 0 \le x_i \le u_i\}$ be an \ILP instance with $\A \in \Z^{r \times c}_{+}$, a non-negative integer matrix with $r$ rows and $c$ columns and $\bm{b} \in \Z_{+} ^ r$, a column vector. 
    In polynomial time, 
    this instance can be reduced to an equivalent instance $\{\A' \bm{y}=\bm{b}':\forall i, \, y_i \in \{0, 1\} \}$ of \ILP where $\A' \in \Z^{r \times c'}_{+}$ is a matrix with $r$ rows, $c' \le c \cdot \max \{u_i + 1\}$ columns and $\norm{\A'}{\infty}=\norm{\A}{\infty}$, and $\bm{b}' \in \Z_{+}^{r}$ is a vector with $\norm{\bm{b}'}{\infty} \le \norm{\bm{b}}{\infty}$.
    \label{lemma:arbitrary_to_binary}
\end{restatable}
\begin{proof}
    For each variable $x_i$, replace each occurrence with $z_i^1 + \ldots + z_i^{u_i}$, where $z_i^1$, \ldots, $z_i^{u_i}$ are $u_i$ new $\{0, 1\}$-variables.
    After this procedure, every constraint
     $$a_1 x_1 + \ldots + a_c x_c = b$$
    is replaced by
    $$a_1 z_1^1 + \ldots + a_1 z_1^{\delta_1} + \ldots + a_c z_c^1 + \ldots + a_c z_c^{\delta_c} = b.$$
    The properties of the resulting instance are straightforward to verify.
\end{proof}
Next, we show that an instance of \ILP with non-negative coefficients and binary variables can be reduced to an instance with binary coefficients.
The proof is similar in spirit to the proof of Lemma 8 in~\cite{knop2020tight}, however we are interested in \ILP instances where variables are only allowed to take values in $\{0, 1\}$, which is the main technical difference.

\begin{restatable}{lemma}{lemmcoerednew}
    Let $\bigl\{ \A   \x=\bm{b}: \forall i, \, x_i \in \{0, 1\} \bigr\}$ be an \ILP instance with $\A \in \Z^{r \times c}_{+}$, a non-negative integer matrix with $r$ rows and $c$ columns and $\bm{b} \in \Z_{+} ^ r$, a column vector. 
    In polynomial time
    , this instance can be reduced to an equivalent instance $\bigl\{\A'  \bm{y}=\bm{b}':\forall i, \, y_i \in \{0, 1\} \bigr\}$ of \ILP where $\A'$ is a $\{0, 1\}$-matrix with $r'= \bo{\log{\norm{\A}{\infty}} \cdot r}$ rows and $c'=\bo{\log{\norm{\A}{\infty}}\cdot (c+\norm{\bm{b}}{\infty})}$ columns, where $\bm{b}' \in \Z_{+}^{r'}$ is a vector with $\norm{\bm{b}'}{\infty}=\bo{\norm{\bm{b}}{\infty}}$.
    \label{lemma:coefficient_reduction_new}
\end{restatable}
\begin{proof}
        Consider the ILP constraints row by row, then every row can be written as $\bm{a}^T \cdot \bm{x}=b$ with $\bm{a} = [a_1, a_2, \dotso, a_c] \in \Z^{ c}_{+} \text{ and } b \in \Z_{+}$ being a row of $\A$ and $\bm{b}$ respectively, where $a_i$ denotes the $i$-th column of $\bm{a}$.
        By choosing $\delta = \ceil{\log(\norm{\A}{\infty}+1)}$, we will have $a_i \leq 2^\delta-1$ and as a result, each $a_i$ can be expressed with $\delta$ bits in binary. 
        Now let $a_i[j] \text{ and } b[j] $ be the $j$-th bit of $a_i$ and $b$ in their binary representations, then $\bm{a}^T \cdot \bm{x}=b$ can be written as:
        \begin{equation}\label{eq:bin_sum}
            \sum_{i=1}^{c}\sum_{j=0}^{\delta - 1} 2^j a_i[j]x_i= \sum_{j=0}^{\delta - 1} 2^j b[j]
        \end{equation}
        where $x_i$ is the $i$-th row of $\bm{x}$.
        In order to achieve $\{0, 1\}$-coefficients, we would like to replace the constraint \eqref{eq:bin_sum} with $\delta$ ``bit-wise'' constraints $\sum_{i=1}^{c} a_i[j]x_i = b[j]$; this is, however, not equivalent to \eqref{eq:bin_sum}, as carry-over might occur.
        Therefore, we additionally introduce carry variables $y_{-1}, y_0, y_1, \dotso, y_{\delta-1}$
        such that $y_{j}$ represents the carry-over obtained from summing \eqref{eq:bin_sum} up to the $j$-th bit; we set $y_{-1}=0$ and $y_{\delta - 1} = \sum_{j \ge \delta} 2^{j-1} b[j]$ to be fixed values in order to unify the first and the last constraint with the rest.
        As a result, \ref{eq:bin_sum} can be replaced with $\delta + 1$ new equations as follows:
        \begin{equation}\label{eq:carry_sum}
            y_{j-1}+\sum_{i=1}^{c} a_i[j]x_i = b[j]+2y_j \text{ for } j \in \{0,1,\dotso,\delta -1 \}
        \end{equation}
        Note that $ a_i[j] \in \{0, 1\}$, so to make (\ref{eq:carry_sum}) a linear constraint with all coefficients being in $\{0, 1\}$, we only need to deal with the $2y_j$ term.
        To this end, for every $y_j$ we introduce two variables $y'_j \text{ and } y''_j$ with constraints $y'_j+y_j = 2^{\ceil{\log(b)}}$ and $y''_j+y_j = 2^{\ceil{\log(b)}}$.
        Now since $y'_j+y''_j=2^{\ceil{\log(b)} +1} - 2y_j$, we can rewrite (\ref{eq:carry_sum}) as:
        \begin{equation}\label{eq:carry_sum_std}
            y_{j-1}+y'_j+y''_j+\sum_{i=1}^{c} a_i[j]x_i = b[j]+2^{\ceil{\log(b)} +1}
        \end{equation}
        Note that $x_i \in \{0,1\}$ while for every $j \in \{0,1,\dotso,\delta\}$ the variables $y_j, y'_j \text{ and } y''_j$ are not necessarily in $\{0,1\}$.
        Since $x_i \in \{0,1\}$ and $a_i[j] \in \{0,1\}$, we can safely assume $y_j \leq c$, and also $y'_j , y''_j \leq 2b \leq 2\norm{\bm{b}}{\infty}$.
        Leaving $H=2\norm{\bm{b}}{\infty}$, we obtain an ILP with $\{0,1\}$ coefficients and $\{0,1\}$ variables by applying the following replacements of variables:
        \begin{itemize}
            \item replacing $y_j$ with the sum of $c$ new variables $z_{1,j}, z_{2,j}, \dotso, z_{c,j}$ with $z_{k,j} \in \{0,1\}$ for $k \in \{1,\dotso ,c\}$.
            \item replacing $y'_j$ and $y''_j$, respectively, with the sum of new variables $z'_{1,j}, \dotso, z'_{H,j}$ and $z''_{1,j},\dotso, z''_{H,j}$ with $z'_{k,j} \in \{0,1\}$ and $z''_{k,j} \in \{0,1\}$ for $k \in \{1,\dotso ,H\}$.
        \end{itemize}
        that is, each constraint of the form (\ref{eq:carry_sum_std}) is replaced with another constraint as follows:
        \begin{equation}
            \sum_{k=1}^{c} z_{k,j-1} + \sum_{k=1}^{H} \bigg( z'_{k,j} + z''_{k,j} \bigg)+ \sum_{i=1}^{r} a_i[j]x_i = b[j]+2^{\ceil{\log(b)} +1}
        \end{equation}
        Note that $b[j]+2^{\ceil{\log(b)} +1} = \bo{\norm{\bm{b}}{\infty}}$.
        As a result, we have an $\ILP$ $\{\A'  \x=\bm{b}': \x \in \{0,1\}^{c'}\}$ with $\A' \in \{0,1\}^{r' \times c'}$ and $\bm{b}' \in \N^{r'}$ where $c' = c + \bigl(\delta + 1 \bigr)\bigl(c+2\norm{\bm{b}}{\infty}\bigr)=\bo{\delta(c+\norm{\bm{b}}{\infty})}$, $r'=(\delta+1)r=\bo{\delta \cdot r}$ and $\norm{\bm{b}'}{\infty} = \bo{\norm{\bm{b}}{\infty}}$. 
\end{proof}
Now we introduce a lemma that allows to reduce a $\{0, 1\}$-coefficient matrix to a matrix with values only in $\{-1,1\}$.
\begin{restatable}{lemma}{lemmbintoplmi}
\label{lemma:binary_to_plus_minus_one}
Let $\{ \A   \x=\bm{b}: x_i \in \{0, 1\} \}$ be an \ILP instance with $\A \in \{0,1\}^{r \times c}$ and $\bm{b} \in \Z_{+} ^ r$.
In polynomial time,
this instance can be reduced to an equivalent instance $\{\A' \bm{y}=\bm{b}':\forall i, \, y_i \in \{0, 1\}\}$ of \ILP where $\A'$ is a $\{-1, 1\}$-matrix with $r'= r+1$ rows and $c'= 2c$ columns and $\bm{b}' \in \Z_{+}^{r'}$ is a vector with $\norm{\bm{b}'}{\infty}= 2\norm{\bm{b}}{\infty}$.
\end{restatable}
\begin{proof}
    Let $\mathbb{J}_{r \times c}$ be a matrix with $r$ rows and $c$ columns such that every entry of this matrix is equal to 1.
    Also, let $\mathbb{1}_r$ be a vector of $r$ entries all equal to 1. 
    Then we have the following equality for any solution $\x \in \{0, 1\}^c$ of $\A   \x = \bm{b}$:
    \begin{equation}
    \label{eq:unit_constraint}
        \bigl( 2\A - \mathbb{J}_{r \times c}  \bigr)   \x = 2\bm{b} - \norm{\x}{1} \cdot  \mathbb{1}_r.
    \end{equation}
    Note that since $x \in \{0,1\}^c$ we have $\norm{\bm{x}}{1}=\sum_{i=1}^{c}x_i$, which is at most $c$.
    By introducing $c$ new variables $z_1, z_2, \dotso, z_c$ with $z_i \in \{0,1\}$ for $i \in \{1,2,\dotso, c\}$ we can rewrite the constraint \cref{eq:unit_constraint} as:
    \begin{equation}
       \begin{split}
            \bigl( 2\A - \mathbb{J}_{r \times c}  \bigr)   \x+ \mathbb{1}_r \cdot \sum_{i=1}^{c} z_i &= 2\bm{b} \\ \nonumber
            \sum_{i=1}^{c} x_i  - \sum_{i=1}^{c} z_i &= 0 
       \end{split}
    \end{equation}
   
    Let $\A'$ be a matrix with $r+1$ rows and $2c$ columns such that the first $r \times c$ entries are equal to $2\A - \mathbb{J}$, the second $r \times c$ entries are all 1, the first $c$ entries of the $(r+1)$-th row are equal to 1 and the second $c$ entries are all -1 (see Figure~\ref{fig:plus_minus_one} for an illustration).
    Let $\bm{b}'$ be a vector with $r+1$ rows such that its first $r$ entries are equal to $2\bm{b}$ and the last entry is 0.
    Also, let $y$ be a vector with $2c$ variables $y_i \in \{0,1\}$ for $i \in \{1, \dotso, 2c\}$. 
    Then the feasibility ILP $\{ \A   \x=\bm{b}: \x \in \{0, 1\}^c \}$ with $\A \in \{0,1\}^{r \times c}$ and $\bm{b} \in \N ^ r$ is equivalent to $\{ \A'   \bm{y}=\bm{b}': \bm{y} \in \{0, 1\}^c \}$ with $\A' \in \{-1,1\}^{(r+1) \times 2c}$ and $\bm{b}' = 2\bm{b}$.
    \begin{figure}[h]
        \centering
        \includegraphics[width=0.5\columnwidth]{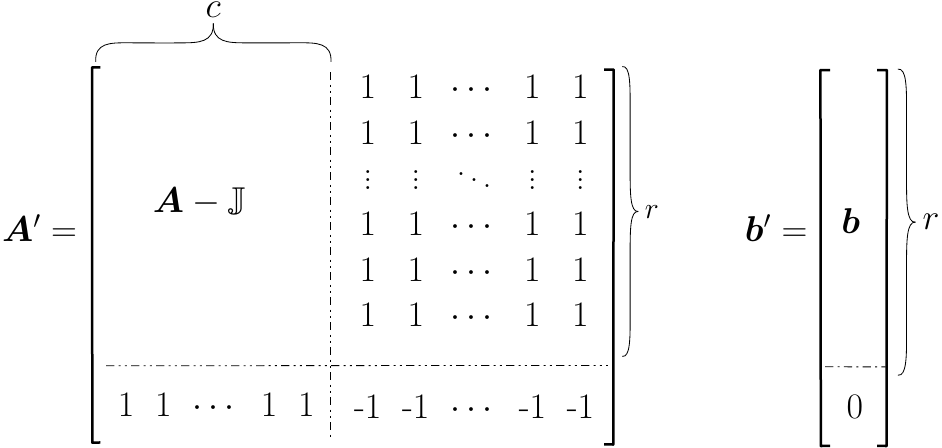}
        \caption{The target instance $(\A', \bm{b}')$ in the construction of Lemma~\ref{lemma:binary_to_plus_minus_one}.} 
        \label{fig:plus_minus_one}
    \end{figure}
\end{proof}
It will be more convenient to work with ILP instances of form $\A   \x \le \bm{b}$, which motivates the following lemma.
\begin{restatable}{lemma}{lemmineq}
    \label{lemma:inequalities}
    Every \ILP instance $\bigl\{ \A   \x \boldsymbol{=} \bm{b}: \forall i, \, x_i \in \{0, 1\} \bigr\}$ with $\A \in \{-1,+1\}^{r \times c}$ and $\bm{b} \in \Z_{+} ^ r$ can be reduced to an equivalent ILP instance $\bigl\{ \A'   \bm{y} \leq \bm{b}': \forall i, \, y_i \in \{0, 1\} \bigr\}$ with $\A \in \{-1,+1\}^{2r \times c}$ and $\bm{b}' \in \Z ^ {2r}$, where $\norm{\bm{b}'}{\infty} = \norm{\bm{b}}{\infty}$.    
\end{restatable}
\begin{proof}
    By changing the constraints to $\A   \x \leq \bm{b}$ and $-\A   \x \leq -\bm{b}$ we get the desired \ILP instance.
\end{proof}
Before we can reduce to \CS itself, we first show a reduction to a slightly more general version of the problem \NUCS, where each string has its own upper bound on the distance to the closest string. For more convenience, we restate the problem.

\begin{definition}[\NUCS] Given a set of $n \in \Z_{+}$ binary strings $S=\set{s_1}{s_n}$ each of length $\ell$ and a distance vector $\bm{d} = \big(d_1, \dotso, d_n \big) \in \Z_{+}^n$, determine whether there exist a binary string $s$ of length $\ell$ such that for all $i \in [n]$:
$$\hd(s, s_i) \leq d_i.$$
\end{definition}

\begin{restatable}{lemma}{lemmnonuniclst}
    \label{lemma:non_uniform_closest_string}
    Every ILP instance $\{ \A   \x \leq \bm{b}: \forall i, \, x_i \in \{0, 1\} \}$ with $\A \in \{-1,+1\}^{r \times c}$ and $\bm{b} \in \Z ^ r$ can be reduced, in polynomial time, to a \NUCS instance with $r$ strings of length $c$, where $d_i \le \bigl(c+\norm{\bm{b}}{\infty} \bigr)$ for each $i \in [n]$.
\end{restatable}
\begin{proof}
    Let $a_i$ and $b_i$ denote the $i$-th row and $i$-th entry of $\A$ and $\bm{b}$ respectively.
    Define $s$ to be the binary string such that its $i$-th bit corresponds to the value of $x_i$ in the ILP, that is $s=[x_1,x_2,\dotso ,x_c]$.
    Also, for each row $a_i$ in $\A$ let $\Tilde{\mathbb{1}}(a_i)=|\bigl\{j \in \{1, \dotso, c\}: a_i[j]=-1 \bigr\}|$ denote the number of $-1$ entries in $a_i$.
    Construct a binary string $s_i \in \{0, 1\}^c$ of length $c$ as follows 
    \[
        s_i[j] = 
        \begin{cases}
              0 & \text{if $a_i[j]=1$}\\
              1 & \text{if $a_i[j]=-1$}
        \end{cases} 
    \]
    and for $j \in \{1,\dotso, c\}$, define variables
    \[
        z_{s_i}^j=
         \begin{cases}
              x_j & \text{if $s_i[j]=0$}\\
              1-x_j & \text{if $s_i[j]=1$}
        \end{cases} 
    \]
     then, the Hamming distance between strings $s$ and $s_i$ can be expressed in terms of variables $z^j_{s_i}$ as follows:
     \[
        \begin{split}
            \hd(s, s_i) &= \sum_{j=1}^{c}z_{s_i}^j = a_i \cdot x+\Tilde{\mathbb{1}}(a_i) \leq b_i + \Tilde{\mathbb{1}}(a_i)
        \end{split}
     \]
    Now let $d_i = b_i + \Tilde{\mathbb{1}}(a_i)$.
    Note that since $\x \in \{0, 1\}^c$ and $a_i \in \{-1, +1\}^c$, if $b_i < 0$ then in order for the ILP instance to be feasible we should have $|b_i| \leq \Tilde{\mathbb{1}}(a_i)$.
    So in a feasible ILP instance, $b_i + \Tilde{\mathbb{1}}(a_i) \geq 0$.
    So from the  ILP instance $\{ \A   \x \leq \bm{b}: x_i \in \{0, 1\} \}$ with $\A \in \{-1,+1\}^{r \times c}$ and $\bm{b} \in \Z ^ r$, we construct a \NUCS instance with the set of strings $\set{s_1}{s_c}$ and the distance vector $\bm{d}=(d_1, \dotso, d_c)$ as described above.
    If for any $i \in \set{1}{c}$ it holds that $d_i < 0$, we get an infeasible instance which indicates that the ILP instance was infeasible in the first place.
    Suppose the ILP instance is feasible, then there is a solution $x \in \{0,1\}^c$ such that $a_i \cdot x \leq b_i$ for all $i \in \set{1}{c}$.
    As a result $ a_i \cdot x+\Tilde{\mathbb{1}}(a_i) = \hd(s, s_i) \leq b_i + \Tilde{\mathbb{1}}(a_i)=d_i$ holds which implies that $s=[x_1,\dotso,x_c]$ is a solution to the \textit{\NUCS} with set of strings $\set{s_1}{s_c}$ and distance vector $\bm{d}=(d_1, \dotso, d_c)$. 
    Now suppose that $s=[x_1,\dotso,x_c]$ is a solution to the \textit{\NUCS} with set of strings $\set{s_1}{s_c}$ and distance vector $\bm{d}=(d_1, \dotso, d_c)$.
    Then $\hd(s, s_i) = a_i \cdot x+\Tilde{\mathbb{1}}(a_i) \leq d_i = b_i + \Tilde{\mathbb{1}}(a_i)$ holds for all $i \in [c]$, which implies $a_i \cdot x \leq b_i$.
    Thus, $x=[x_1,\dotso, x_c]$ is a solution to the original ILP instance.
    So if $d_i < 0$ for any $i \in [c]$, the ILP instance is feasible.
\end{proof}
Finally, we show a reduction from \NUCS to \CS.
\begin{restatable}{lemma}{lemmclostr}
\label{lemma:closest_string}
    Every \NUCS instance with strings $S=\set{s_1}{s_n}$ of length $\ell \in \N$ and distance vector $\bm{d}=(d_1, \dotso, d_n)$ can be reduced to a \CS instance with $2n$ strings $S'=\set{s'_1}{s'_{2n}}$ and distance $d = \max_{i=1}^{n} d_i$ in time $\bo{n \cdot \ell}$, where length of the constructed strings is at most $\ell + 2n \cdot d$.    
\end{restatable}
\begin{proof}
    For any binary string $s$ of length $\ell$ and $i,j \in [\ell]$ with $i<j$, let $s[i,j]$ be the substring obtained from $s$ that is restricted to indices $i$ through $j$, inclusive.
    We write $s[i,j]=\overline{0}$ (and $s[i,j]=\overline{1}$) to set the bits from index $i$ to index $j$ (inclusive) in the string $s$ to 0 (and 1 respectively).
    Without loss of generality assume that $d_1 \leq d_2 \leq \dotso \leq d_n$, then $d=\max_{i=1}^{n} d_i=d_n$.
    Also define $\Delta_i = d - d_i$, $p_i=\ell+2\sum_{j=1}^{i-1} \Delta_{j}$ and $\Delta = \sum_{i=1}^{n} \Delta_i$.
    For each string $s_i$ in the \NUCS instance, we construct two corresponding strings $s^1_i$ and $s^2_i$ each of length $\ell + 2\Delta$ in our \CS instance, as follows:
    \begin{itemize}
        \item Set the first $\ell$ bits of $s^1_i$ and $s^2_i$ to $s_i$, \textit{i.e.} $s^1_i\big[1,\ell \big]=s^2_i\big[1,\ell \big]=s_i$.
        \item Among the remaining $2\Delta$ bits of $s^1_i$, set the $\Delta_i$ bits from index $(p_i+1)$ through $(p_i + \Delta_i)$ to 1 and the rest to 0. 
        \textit{i.e.} $s^1_i \big[\ell+1, p_i \big]=\overline{0}, \,\, s^1_i \big[p_i+1, p_i + \Delta_i \big]=\overline{1} \text{ and } s^1_i \big[p_i + \Delta_i+1, \ell+2\Delta \big]=\overline{0}$.
        \item For $s^2_i$, set the $\Delta_i$ bits from index $(p_i+\Delta_i+1)$ through $(p_{i+1})$ to 1.
        \textit{i.e.} $s^2_i \big[\ell+1, p_i+\Delta_i \big]=\overline{0}, \,\, s^2_i \big[p_i+\Delta_i+1, p_{i+1} \big]=\overline{1} \text{ and } s^2_i\big[p_{i+1}+1, \ell+2\Delta \big]=\overline{0}$.
    \end{itemize}
the resulting strings are illustrated in Figure(\ref{fig:stringConcatenation}).
Note that $\Delta_n = d - D_n = 0$.\\
 \begin{figure}
     \centering
     \includegraphics[width=0.7\columnwidth]{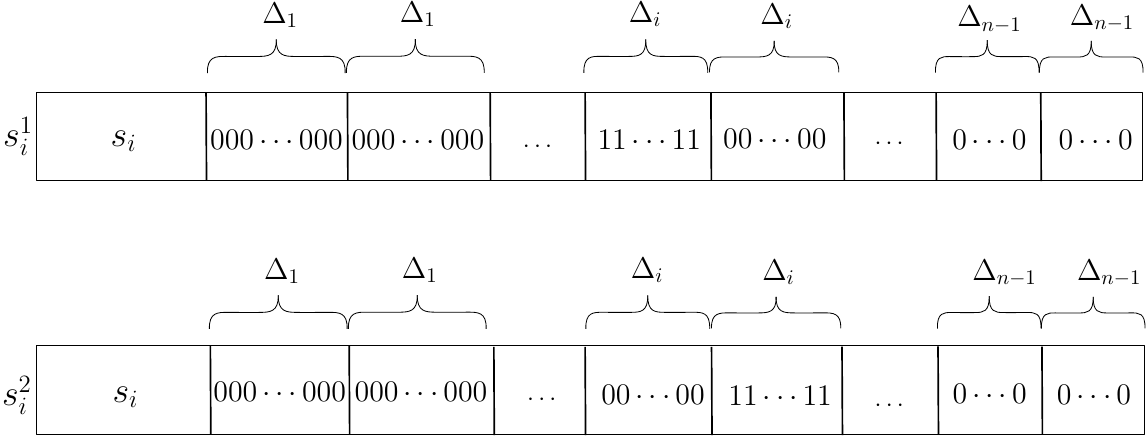}
     \caption{Construction of strings in the Closest String instance, Lemma~\ref{lemma:closest_string}.}
     \label{fig:stringConcatenation}
 \end{figure}
Define $S'=\bigcup_{i=1}^{n} \bigl\{s_i^1, s_i^2\bigr\}$ to be the set of all $2n$ strings we have constructed.
Let $s$ be a solution to the \NUCS instance.
We construct a string $s'$ of length $\ell+2\Delta$ that is obtained by appending $2\Delta$ zeros to $s$.
Obviously $s'$ is a valid solution to the \CS instance, since for every $i \in [n]$ and $j \in \{1,2\}$, we have:
\[
\begin{split}
        \hd \bigl(s',s^j_i \bigr)&=\hd \bigl(s'[1, \ell],s^j_i[1, \ell]\bigr)
        +\hd\bigl(s'[\ell+1, \ell + 2\Delta],s^j_i[\ell+1, \ell + 2\Delta]\bigr) \\
        &=  \hd \bigl(s,s^j_i[1, \ell]\bigr) + \Delta_i \leq d_i + \Delta_i = d
\end{split}
\]
the last inequality holds since $s^j_i[1, \ell] = s_i $. Thus the binary string $s'$ has Hamming distance of at most $d$ to all the strings in the set $S'$.
Now let $s'$ be a valid solution to the \CS instance.
Define sets $I, \overline{I} \subseteq \N$ as $I = \bigl\{p_{i}+1, \dotso, p_{i+1} \bigr\}$ and $ \overline{I} = \bigl\{\ell+1, \dotso, \ell + 2\Delta \bigr\} \!\setminus\! I $.
For $i \in [n]$ and $j \in \{1,2\}$, we partition $ \hd\bigl(s',s_i^j\bigr)$  as follows:
\begin{equation}
    \label{Eq:distancePartition}
    \setlength{\jot}{5pt} 
    \begin{split}
        \hd\bigl(s',s_i^j\bigr) = \hd\bigl(s'\big[1, \ell \big],s_i^j \big[1, \ell \big]\bigr) \notag
                              + \hd\bigl(s' \big[\overline{I}\big],s_i^j \big[\overline{I} \big]\bigr) \notag
                              + \hd\bigl(s' \big[I \big],s_i^j \big[I \big]\bigr)
    \end{split}
\end{equation}

According to the construction, the first $l$ bits of $s^1_i$ and $s^2_i$ have the same values as $s_i$, therefore $$\hd\bigl(s'[1, \ell],s_i^j[1, \ell]\bigr)=\hd\bigl(s'[1, \ell],s_i\bigr).$$  
Moreover, among the remaining $2\Delta$ bits, it holds that $s_i^1[\overline{I}] = s_i^2[\overline{I}]$ and $s_i^1[I] = \overline{s_i^2[I]}$  since both $s^1_i$ and $s^2_i$ are zero on indices indicated by the set $\overline{I}$ and are complement of each other on indices indicated by the set $I$. 
Thus, $\hd\bigl(s' \big[\overline{I}\big],s_i^{1} \big[\overline{I} \big]\bigr) = \hd\bigl(s' \big[\overline{I}\big],s_i^{2} \big[\overline{I} \big]\bigr)$ and $\hd\bigl(s' \big[I\big],s_i^{1} \big[I \big]\bigr) = 2\Delta_i - \hd\bigl(s' \big[I\big],s_i^{2} \big[I \big]\bigr)$.
As a result, using \cref{Eq:distancePartition}, we obtain
\begin{equation*}
\label{Eq:sumDistance}
    \begin{split}
         \sum_{j=1}^{2} \hd\bigl(s',s_i^j\bigr) &= 2\hd\bigl(s'[1, \ell],s_i\bigr) + 
        2\hd\bigl(s'[\overline{I}],s_i^1[\overline{I}]\bigr) + 2\Delta_i \leq 2d \\
        &\Rightarrow \hd\bigl(s'[1, \ell],s_i\bigr) + 
        \hd\bigl(s'[\overline{I}],s_i^1[\overline{I}]\bigr) \leq d - \Delta_i = d_i
    \end{split}
\end{equation*}
and since $\hd\bigl(s'[\overline{I}],s_i^1[\overline{I}]\bigr) \geq 0$, we have that $\hd\bigl(s'[1, \ell],s_i\bigr) \leq d_i$.
So the string $s=s'[1, \ell]$ (the first $l$ bits of $s'$) is a valid solution to the original instance of \NUCS.
\end{proof}
With the reductions above at hand, we can conclude with the proof of Theorem~\ref{thm:hardness} as outlined below.
\begin{proof}[Proof of Theorem~\ref{thm:hardness}]
    We start by applying Lemma~\ref{lemma:smallerdomain}, to obtain an equivalent \ILP instance where the coefficients, constraints and variable domains are all non-negative integers.
    Next, by applying Lemma~\ref{lemma:arbitrary_to_binary}, we get an equivalent instance of \ILP with binary variables.
    Then, with the help of Lemma~\ref{lemma:coefficient_reduction_new}, we get that also the coefficients in the constraints are binary.
    After applying Lemma~\ref{lemma:binary_to_plus_minus_one}, we get that all coefficients are either $-1$ or $+1$, and additionally that the ILP instance is of the form $\A   \x \le \bm{b}$ by applying Lemma~\ref{lemma:inequalities}.
    Finally, we reduce to \NUCS via Lemma~\ref{lemma:non_uniform_closest_string}, and then to \CS via Lemma~\ref{lemma:closest_string}.
    
    We now argue about the size of the constructed instance.
    The number of rows is increased by a factor of $\bo{\log\norm{\A}{\infty}}$ by the reduction of Lemma~\ref{lemma:coefficient_reduction_new}, and by at most a constant factor in all other reductions.
    Therefore, $n = \bo{\log \norm{\A}{\infty} \cdot r}$.
    The number of columns is increased by the original number of rows in Lemma~\ref{lemma:smallerdomain}, by a factor of $\delta$ in Lemma~\ref{lemma:arbitrary_to_binary}, where $\delta = \max u_i - \ell_{i} + 1$, and then $\bo{c \cdot \delta \cdot \norm{\A}{\infty}}$ additional columns are added in Lemma~\ref{lemma:coefficient_reduction_new}, along with an additional factor of $\bo{\log\norm{\A}{\infty}}$.
    Lemma~\ref{lemma:binary_to_plus_minus_one} only increases the number of columns by at most a constant factor, and Lemmata~\ref{lemma:inequalities}, \ref{lemma:non_uniform_closest_string} do not change the number of columns.
    Finally, in Lemma~\ref{lemma:closest_string}, $d$ is set to at most $\bo{\norm{\bm{A}}{\infty} \cdot \log{\norm{\A}{\infty}}\cdot c \cdot \delta}$ and additional $2n\cdot d$ columns are attached, where $n = \bo{\log \norm{\A}{\infty} \cdot r}$.
Therefore, the final length of the strings $\ell$ is at most $\bo{\norm{\A}{\infty} \cdot \log^2{\norm{\A}{\infty}}\cdot r \cdot c \cdot \delta}$.
\end{proof}

As mentioned in the previous section, \ILP admits an algorithm with running time $(r\Delta)^{(r + 1)^2} \cdot \poly(r c)$~\citep{eisenbrand2019proximity}, where $\Delta = \norm{\A}{\infty}$.
On the other hand, a version of \ILP without upper bounds on the variables can be solved in time $\bo{c \cdot (r\Delta)^{2r} \cdot \norm{\bm{b}}{1}^2}$~\citep{eisenbrand2019proximity}, and this is tight under the ETH~\citep{knop2020tight}.
This raises a natural question whether the gap in the running time between the two versions is necessary.
As \CS admits a straightforward reduction to \ILP, it also makes sense to ask whether running time better than $n^{\bo{n^2}} \cdot \poly(n \ell)$ can be achieved for \CS.
From Theorem~\ref{thm:hardness}, it follows that an improved \CS algorithm would automatically improve the best-known running time for \ILP (with upper bounds).
Formally, we prove Theorem~\ref{thm:ilp}, as stated in~\Cref{sec:intro}.
\thmilp*
\begin{proof}
    We construct a \CS instance following Theorem~\ref{thm:hardness}, and then by applying the assumed algorithm for \CS on the obtained instance, we obtain an algorithm for the original \ILP with running time:
    $$2^{\bo{n^{1 + \alpha}}} \cdot \poly(n \ell),$$
    where $n = \bo{\log{\norm{\A}{\infty}} \cdot r}$, $\ell = \bo{\norm{A}{\infty} \cdot \log^2{\norm{A}{\infty}}\cdot r \cdot c \cdot \delta}$.
    This can be expressed as
    \begin{align*}
       2^{\bo{n^{1 + \alpha}}} \poly(n \ell) &=
       2^{\bo{(\log{\norm{\A}{\infty}} \cdot r)^{1 + \alpha}}} \cdot (\norm{\A}{\infty} \cdot r \cdot c \cdot \delta)^{\bo{1}}\\ 
         &=2^{\bo{r^{1 + \alpha + o(1)}}} \cdot r^{\bo{1}} \cdot c^{\bo{1}} \cdot 2^{\bo{r}}\\
         &=2^{\bo{(r^{1 + \alpha + o(1)})+1}} \cdot \poly(rc). \\
         &=2^{\bo{r^{1 + \alpha + o(1)}}} \cdot \poly(rc).
    \end{align*}
\end{proof}
From Theorem 1 in~\cite{knop2020tight}, and the chain of reductions in the proof of Theorem~\ref{thm:hardness}, we can also get the following hardness result for \CS, that at least the running time of $2^{\bo{n \log n}} \cdot \poly(n \ell)$ is necessary, similarly to $\ILP$ with unbounded variables.
\begin{corollary}
    Assuming ETH, there is no algorithm that solves \CS in time $2^{o(n \log n)} \cdot \poly(n \ell)$.
\end{corollary}

    \section{Vertex Cover}
     \label{sec:vc_fpt}
    In this section, we describe an FPT algorithm for \kCME parameterized by the vertex cover number of the incidence graph, $G_{\M}$, derived from mask matrix $\M$.
More formally, we prove the following:
\thmvc*
\begin{proof}
Let  $\vxc$ denote the minimum vertex cover of $G_{\M}$, which can be computed in time $\bo{1.2738^{\vc(G_{\M})} + (\vc(G_{\M}) \cdot nm)}$ by the state-of-the-art algorithm of~\cite{ChenKX06}.
We define the sets $\rvc = (\vxc \cap R_{\M})$ and $\cvc = (\vxc \cap  C_{\M})$, to represent the rows and coordinates in $\M$ that are part of the vertex cover $\vxc$.
Additionally, we define $\overline{\cvc} = C_{\M} \setminus \cvc$, and $\overline{\rvc} = \big\{ r \in (R_{\M} \setminus \rvc) : \A[r][\cvc] \neq \overline{0}\big\}$ as the set of rows outside $\rvc$ whose substring induced by the coordinates in $\cvc$ contains at least one non-zero entry.
Remember that, by the definition of $G_{\M}$, for every $r \in \overline{\rvc}$ and $c \in \overline{\cvc}$, it follows that $\M[r][c] = 0$.
To see this, assume there exist $r_0 \in \overline{\rvc} , c_0 \notin \overline{\cvc}$ such that $\M[r_0][c_0] = 1$.
Then, the edge $(r_0, c_0)$ exists in $G_{\M}$ that is not covered by $\vxc$, as neither $r_0$ nor $c_0$ belong to $\vxc$, which contradicts that $\vxc$ is a vertex cover of $G_{\M}$.

Consequently, the rows and coordinates in $(\overline{\rvc} \cap \overline{\cvc})$ do not contribute to the radius of any cluster, since $\M[\overline{\rvc}][\overline{\cvc}] = 0$ which means that $\A[\overline{\rvc}][\overline{\cvc}] = ?$.
Therefore,  we only need to focus on the entries in $(\rvc \cap \cvc)$ for cluster assignments.
For further clarification, refer to \cref{fig:vertex_cover_graph}.
The rows contained in $\rvc$ will be referred to as \emph{long rows}, and the rows within $\overline{\rvc}$ will be called \emph{short rows}, as they contain values solely along the coordinates within $\cvc$. See \Cref{fig:vertex_cover_graph} for an illustration of the instance structure; rows and coordinates may need to be reordered to reflect this structure, but this does not affect the problem's objective or the algorithm.
\begin{figure}[h]
    \centering
    \includegraphics[width=0.28\textwidth]{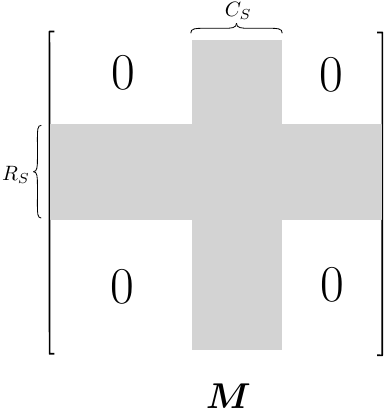}
    \caption{The mask matrix $\M$. The non-zero entries are distributed only within $\rvc$ and $\cvc$. For every $r \in \overline{\rvc}$ and $c \in \overline{\cvc}$ it holds that $\M[r][c]=0$.}
    \label{fig:vertex_cover_graph}
\end{figure}

We also define the mappings: 
\begin{itemize}
    \item $\parti: \rvc \rightarrow [k]$,
    \item $\cent: [k] \times \cvc \rightarrow \{0, 1\}$
\end{itemize}
which we refer to as a \emph{partial cluster assignment} and a \emph{partial center assignment}, respectively.
Intuitively, $\parti$ assigns each long row $r \in \rvc$ to one of the $k$ clusters, and $\cent$ assigns values to the centers of \emph{all} clusters, only along the coordinates in $\cvc$.
We call a pair $(\parti, \cent)$, a \emph{partial assignment} and say it is \emph{valid} if, for every $r \in \rvc$, the condition $\hd \bigl(\A[r][\cvc], \cent \bigl[\parti[r], \cvc \bigr]\bigr) \leq d$ holds.
Let $P$ denote the set of all possible partial assignments, and for a valid $(\parti, \cent) \in P$, define $J = \bigcup_{r \in \rvc} \parti[r]$, as the set of clusters assigned to the long rows by $\parti$. 
For a cluster $j \in J$ with cardinality $p_j$, let $T_j = \big\{r_{j_1}, r_{j_2}, \cdots, r_{j_{p_j}}\big\} \subseteq \rvc $ be the set of long rows that $\parti$ assigns to cluster $j$.

Note that, in order to ensure a correct cluster assignment, we must verify that the distance between each long row and its cluster center along the coordinates in $C_{\M}$ does not exceed $d$, and that the distance between each short row and its cluster center along the coordinates in $\cvc$ also remains within $d$.
\\
With the necessary definitions set, we now proceed to outline Algorithm~\ref{alg:vertex_cover_fpt}:
\\
First, we start by obtaining the minimum vertex cover of $G_{\M}$. 
Then we establish a valid partial assignment $(\parti, \cent)$ for the \emph{long} rows along coordinates in $\cvc$ \textit{(lines 1-4)}.
At this point, each long row has some distance to its corresponding cluster center along the coordinates in $\cvc$, that is $\hd(r[C_S], \cent[\parti[r]])$. 
Consequently, the remaining distance that each long row $r_{j_i} \in T_j$ can have to its respective cluster center along coordinates in $\overline{\cvc}$ is bounded by $d'_{j_i} = d - d_H \bigl(\A[r_{j_i}][\cvc], \cent[j, \cvc] \bigr)$ \textit{(lines 5-11)}.
\\
As the next step, for the clusters that have been assigned to the \emph{long} rows, we try to determine if there exists a valid center assignment along the remaining coordinates, $\overline{\cvc}$.  To do so, we solve a binary $\NUCS$ instance (denoted as \textit{NUCS} on line 14) for each cluster $j \in J$, with the distance vector $\bm{d}'_j=(d'_{j_1}, \dotso, d'_{j_{p_j}})$, and the strings $r'_{i_j} = \A[r_{i_j}][\overline{\cvc}]$, for $i \in [p_j]$  \textit{(lines 12-16)}. At this stage, we have a valid cluster assignment for the long rows, and the remaining task is to assign the short rows to appropriate clusters.
\\
Finally, we decide whether a valid cluster assignment is also possible for the \emph{short} rows. So, we assign each $\overline{r} \in \overline{\rvc}$, to the smallest $j \in [k]$ such that $\hd \bigl(\A[\overline{r}][\cvc], \cent[j, \cvc] \bigr) \leq d$ \textit{(lines 17-24)}.  If at this step, we can assign all the \emph{short} rows ($ \overline{\rvc}$) to some cluster center defined by $\cent$, then the given $\kCME$ instance is a feasible \textit{(lines 25-26)}. Otherwise, we repeat the process for another valid partial assignment.\\
For correctness, first assume there is a solution, we show that our algorithm correctly finds it.
Let $\parti^{*}:R_{\M} \rightarrow [k]$ be the solution cluster assignment and $\cent^{*}:[k] \times C_{\M} \rightarrow \{0, 1\}$ be the solution center assignment.
Then for each cluster $i \in [k]$ and each long row $r \in \rvc$ assigned to it, that is $\parti^{*}[r] = i$, it holds that 
\begin{align*}
    \hd(\A[r],\cent^{*}[i]) = \hd \bigl(\A[r][\cvc],\cent^{*}[i][\cvc]\bigr)+ \hd \bigl(\A[r][\overline{\cvc}],\cent^{*}[i][\overline{\cvc}]\bigr) \leq d
\end{align*}
Since we consider all the possible valid partial assignments for long rows in our algorithm, $\cent^{*}[\parti^{*}[r]][\cvc]$ is captured by at least one of the partial assignments $(\parti, \cent) \in P$.
In other words, there is at least one partial assignments $(\parti, \cent) \in P$ such that $\parti = \parti^{*}|_{R_S}$ and $\cent = \cent^{*}|_{C_S}$, where $|$ denotes restriction to a set.
Thus, by definition, it holds that $\hd(\A[r][\overline{\cvc}],\cent^{*}[\parti^{*}[r]][\overline{\cvc}]) \leq d'_r$.
Consequently, the $\NUCS$ instance for cluster $i$ and the long rows assigned to it, will correctly output feasible.
As a result, for the long rows, we correctly output that a $k$-cluster with radius at most $d$ is feasible.
For the short rows, note that they have non-missing entries only along coordinates in $\cvc$, so for each short row $\overline{r} \in \rvc$ assigned to cluster $i \in [k]$, that is $\parti^{*}[\overline{r}] = i$, we have:
\begin{align*}
    \hd(\A[\overline{r}],\cent^{*}[i]) &= \hd \bigl(\A[\overline{r}][\cvc],\cent^{*}[i, \cvc]\bigr) \\
                                       &= \hd \bigl(\A[\overline{r}][\cvc],\cent[i, \cvc]\bigr) \leq d.
\end{align*}
Hence, $\overline{r}$ will at least be assigned to cluster $i$.
Therefore if there is a solution to this instance of $\kCME$, then our algorithm will successfully output feasible.
The other direction can similarly be showed.\\
As of the running time, in the first step, we obtain the minimum vertex cover of $G_{\M}$ in time $\bo{1.2738^{\vc(G_{\M})} + (\vc(G_{\M}) \cdot nm)}$ using the algorithm described in~\cite{ChenKX06}.
For the following steps, note that there are at most ${k}^{|\rvc |}=k^{\bo{\vc{(G_{\M})}}}$ possible mappings for $\parti$ and $2^{(|\cvc| \cdot |J|)} = 2^{\bo{\vc(G_{\M}) \cdot k}}$ for $\cent$. 
The $\NUCS$ instances are solved via ILP formulation described in \cref{sec:CS_to_ILP} in time $m \cdot \bo{\vc(G_{\M})}^{\bigl(\vc(G_{\M})+1\bigr)^2+1} \cdot \log^2{\vc(G_{\M})}$, according to \cref{co:binary_closest_string_time}.
Note that because the distance of each short row $\overline{r} \in {\overline{R_{\vxc}}}$ to a cluster center can be checked in time $(k' \cdot |\cvc|) = \bo{k \cdot \vc(G_{\M})}$, the final step can be computed in time $\bo{m \cdot k \cdot \vc(G_{\M}) \cdot 2^{(\vc(G_{\M}) \cdot k)}}$.
In total, the running time of the algorithm is dominated by $2^{\bo{k \cdot \vc(G_{\M}) + \vc(G_{\M})^2 \cdot\log{\vc(G_{\M})}}} \poly(nm).$

\begin{algorithm}[h]
\caption{\kCME parameterized by $\vc(G_{\M})$} \label{alg:vertex_cover_fpt}
\KwIn{$k, d \in \N$, $\A \in \{0,1,?\}^{n \times m}$ and $\M \in \{0, 1\}^{n \times m}$}
\KwOut{Yes/No}
$S \gets \text{Minimum Vertex Cover of } G_{\M}$\\
$P \gets \big\{(\parti, \cent)\in ( \rvc \rightarrow [k]) \times ([k] \times \cvc \rightarrow \{0, 1\})\big\}$\\
\For{$(\parti, \cent) \in P $}{
    \If{$\text{isValid}(\parti, \cent)$}{
        $J \gets \bigcup_{r \in \rvc} \parti[r]$\\
        $\text{soln} \gets true$ \\
        \For{$j \in J$}{
            $T_j \gets \big\{r_{j_1}, r_{j_2}, \cdots, r_{j_{p_j}}\big\} $\\
            
            \For {$r_{j_i} \in T_j$}{
                $d'_{j_i} = d - \hd \bigl(\A[r_{j_i}][\cvc], \cent[j, \cvc] \bigr)$\\
                $r'_{j_i} = \A[r_{j_i}][\overline{\cvc}]$\\
            }
            $d' \gets (d'_{j_1},\dotso, d'_{j_{p_j}})$\\
            $r' \gets (r'_{j_1}, \dotso, r'_{j_{p_j}})$\\
            $\text{soln} \gets \text{soln} \wedge NUCS \bigl(k,d',r'  \bigr)$\\
            \If{$!\text{soln}$} {break}
        }
        \If{$\text{soln}$}{
            $\overline{\rvc} = \big\{ r \in (R_{\M} \!\setminus\! \rvc) : \A[r][\cvc] \neq \overline{0}\big\}$\\
            $counter \gets 0 $\\
            \For{$r \in \overline{\rvc}$}{
                \For{$j \in [k] $}{
                    \If{$\hd \bigl(\A[r][\cvc], \cent[j, \cvc] \bigr) \leq d$}{
                        $counter++$\\
                        break
                    }
                }
            }
            \If{$counter = |\overline{\rvc}|$}{ 
            \Return YES
            }
        }
    }
}
\Return NO
\end{algorithm}
\end{proof}
\vspace{-.7cm}
    \section{Treewidth}
     \label{sec:tr_fpt}
    This section is dedicated to a fixed-parameter algorithm for the problem parameterized by the treewidth of the incidence graph $G_{\M}$, the number of clusters $k$ and the maximum permissible radius of the cluster, $d$. We restate the formal result next for convenience.

\thmtw*

First, we briefly sketch the intuition of our approach. Informally, the fact that $G_{\M}$ has treewidth at most $t$ means that the graph can be constructed in a tree-like fashion, where at each point only a vertex subset of size at most $t$ is ``active''. This small subset is called a $\emph{bag}$, and in particular is a separator for the graph: there are no edges between the ``past'', already constructed part of the graph and the ``future'', not encountered yet part of the graph; all connections between these parts are via the bag itself. In terms of the \kCME instance, this means that for each ``past'' row, all its entries that correspond to ``future'' columns are missing (as the respective entry of the mask matrix $\M$ has to be $0$), and the same holds for ``past'' columns and ``future'' rows.

Our algorithm performs dynamic programming over this decomposition, supporting a collection of records for the current bag. For the rows of the current bag, we store the partition of the rows into clusters, and additionally for each row the distance to its center vector among the already encountered columns. For each cluster and for each column of the bag, we store the value of the cluster center in this column. These three characteristics (called together a \emph{fragment}) act as a ``trace'' of a potential solution on the current bag. In our DP table, we store whether there exists a partial solution for each choice of the fragment. Because of the separation property explained above, only knowing the fragments is sufficient for computing the records for every possible update on the bag. The claimed running time follows from upper-bounding the number of possible fragments for a bag of size at most $t$.

Now we proceed with the formal proof.
We first recall the definition of a nice tree decomposition, which is a standard structure for performing dynamic programming in the setting of bounded treewidth.

\subparagraph*{\textbf{Nice Tree Decomposition.}}
A \emph{nice tree decomposition} of a graph $G=(V, E)$ is a pair $(\mathcal{T, X})$ where $\mathcal{T}$ is a rooted tree at node $r$ and $\mathcal{X}$ is a mapping that assigns to each node $t \in \mathcal{T}$ a set $X(t) \subseteq V(G)$, referred to as the bag of at node $t$. A nice tree decomposition satisfies the following properties for each $t \in \mathcal{T}$:
\begin{enumerate}
	\item $X(r) = \emptyset$ and for every leaf $l$, it holds that $|X(l)| = 1$. \label{tw:leafs}
	\item For every $uv \in E$, there is a node $t \in \mathcal{T}$ such that both $u \in X(t)$ and $v \in X(t)$ hold. \label{tw:edges}
	\item For every $v \in V$, the set of nodes $t \in T$ such that $v \in X(t)$, induces a connected subtree of $\mathcal{T}$. \label{tw:connected}
	\item There are only three kinds of nodes (aside from root and the leafs) in $\mathcal{T}$:
	\begin{enumerate}
		\item \textbf{Introduce Node:} An introduce node $t$ has exactly one child $t'$ such that there is a vertex $v \notin X(t')$ satisfying $X(t) = \{v\} \cup X(t')$. We call $v$, the introduced vertex.
		\item \textbf{Forget Node:} A forget node $t$ has exactly one child $t'$ such that there is a vertex $v \notin X(t)$ satisfying $X(t) = X(t') \!\setminus\! \{v\}$. We call $v$, the forgotten vertex.
		\item \textbf{Join Node:} A join node $t$ has exactly two children $t_1$ and $t_2$ such that $X(t)=X(t_1)=X(t_2)$.
	\end{enumerate}
\end{enumerate}
Note that by properties \ref{tw:edges} and \ref{tw:connected}, while traversing from leaves to the root, a vertex $v \in V(G)$ can not be introduced again after it has already been forgotten. Otherwise the subtree of $\mathcal{T}$ induced by the nodes whose bags contain $v$, will be disconnected.
The \emph{width} of a nice tree decomposition $(\mathcal{T, X})$ is defined as $\max_{t \in \mathcal{T}} \bigl(|X(t)|-1 \bigr)$ and the \emph{treewidth} of a graph $G$ is defined to be the smallest width of a nice tree decomposition of $G$ and is denoted by tw$(G)$. \\
One can use fixed-parameter algorithms described in \cite{bodlaender1993linear} and \cite{kloks1994treewidth} to obtain a nice tree decomposition with the optimal width and linearly many nodes.
However for a better running time, fixed-parameter approximation algorithms are often used. Specifically, in this paper we apply the 5-approximation algorithm established in \cite{bodlaender2016c} to obtain a nice tree decomposition of $G_{\M}$ with width $q \leq 5\cdot\tw(G_{\M})$ in time $2^{\bo{\tw(G_{\M})}}(n+m)$.\\
So let $(\mathcal{T, X})$ be the nice tree decomposition of $G_I$ rooted at $r$ with treewidth $q \le 5\cdot \tw($G$)$.
Remember that $V(G_I)$ corresponds to the rows and coordinates of the mask matrix $M$, so for a node $t \in \mathcal{T}$, we denote the rows and coordinates in $X(t)$ by $R_t$ and $C_t$ respectively.
At each note $t$, we write $n_{R_t}$ and $n_{C_t}$ to denote the cardinality of the sets $R_t$ and $C_t$, respectively.
Also, we denote by $\mathcal{T}_t$, the subtree of $\mathcal{T}$ rooted at $t$ and define the set of all bags in $\mathcal{T}_t$ by $\down{X_t}$, that is $\down{X_t}= \bigcup_{t' \in \mathcal{T}_t}X(t')$. Moreover we extend the notation and denote the set of rows and coordinates in $\mathcal{T}_t$ respectively by $\down{R}_t$ and $\down{C}_t$. \\

We now recall the statement of \Cref{thm:tw} and start with formally explaining the dynamic programming approach that starts from the leaves and traverses toward the root \( r \), computing and storing the relevant records at each node \( t \in \mathcal{T} \). Once the records at \( r \) are computed, they are used to derive the correct solution. Intuitively, at each node \( t \in \mathcal{T} \), these records will store: a partitioning of rows in \( X(t) \) into clusters, the cluster centers limited to the coordinates present in \( X(t) \), and the potential distances between the rows in the bag and their cluster centers along all the coordinates visited up to that node.
\begin{proof}
We continue to formally explain the records and how the dynamic programming proceeds. At every node $t$, we define the following mappings:
\begin{itemize} 
	\item $\parti: R_t \rightarrow [k]$,
	\item $\cent: [k] \times C_t \rightarrow \{0,1\}$,
	\item $\dist: R_t \rightarrow [d]$,
\end{itemize} 
and call a triple $(\parti, \cent, \dist)$ a \emph{fragment} in $t$. Intuitively, at node $t$, $\parti $ defines a (partial) clustering of the rows within the bag, $\cent$ assigns value to all cluster centers along the coordinates in the bag and $\dist$ considers, for each row $r_i$ in the bag, all possible Hamming distances to its corresponding cluster center, along the coordinates in $\down{C}_t$. We say $\dist$ is valid if for all $j \in R_{t}$ it holds that $0 \leq d_j \leq d $\\
Moreover, let $\mathcal{B}_t$ be the set of all binary matrices with row labels in $\down{R}_t$ and coordinate labels in $\down{C}_t$. We say $(\parti, \cent)$ is a \emph{partial fragment} of $B_t \in \mathcal{B}_t$ at $t$, if there is a cluster assignment $\phi$ with respect to $B_t$ such that:
\begin{itemize}
	\item $\parti = \phi |_{R_t}$,
	\item For every $c \in C_t, \text{ and } r \in \down{R}_t: \B_t[r][c] = \cent[\phi(r), c] $. 
\end{itemize}
Recall that the existence of a cluster assignment implies, in particular, that $\B_t$ has at most $k$ distinct rows.
For a mapping $f: X \rightarrow Y$ and a set $S$, we use the notation $f|_{S}$ to denote the restriction of $f$ to the elements of $S$.
Let $P(t)$ be the set of all fragments at $t$, then our dynamic programming records, $\Dp_t:P(t) \rightarrow \{0, 1\}$ will be a mapping from each fragment at $t$ to a number in $\{0, 1\}$, as follows.
For a fragment $(\cent, \parti, \dist)$, with $\dist = (d_1, d_2,\cdots,d_{n_{R_t}})$, we set $\Dp_t \bigl(\parti, \cent, \dist\bigr)  =  1$ if there is a $B_t \in \mathcal{B}_t$ such that:
\begin{enumerate}[label=\alph*)]
	\item $(\parti, \cent)$ is a partial fragment of $\B_t$ at $t$, \label{dp:a}
	\item $\forall r \in R_t: \; \hd \bigl(\A[r][\down{C}_t], \B_t[r][\down{C}_t] \bigr) = \dist[r]$, \label{dp:b}
	\item $\forall r \in \down{R}_t \!\setminus\! R_t: \; \hd \bigl(\A[r][\down{C}_t], \B_t[r][\down{C}_t] \bigr) \leq d$, \label{dp:c}
\end{enumerate}
and we say $\dist$ \emph{fits} $\B_t$ and $\B_t$ \emph{confirms} $\Dp_t \bigl(\parti, \cent, \dist\bigr) = 1$.
Observe that, according to the definition of the nice tree decomposition, at the root node $r$, we have $X(r) = R_r = C_r = \emptyset$, which implies that $P(r) = {(\emptyset, \emptyset, \emptyset)}$. 
Furthermore, $\down{R}_r = R_{\M} = R_{\A}$ and $\down{C}_r = C_{\M} = C_{\A}$. 
Therefore, by properties \ref{dp:a} and \ref{dp:c} of the dynamic programming, $\Dp_r \bigl(\emptyset, \emptyset, \emptyset \bigr)=1$ indicates that there exists a matrix $B_r$ defined over rows $R_{\A}$ and coordinates $C_{\A}$ with at most $k$ distinct rows, such that for all $r \in R_{\A}$ it holds that $\hd\bigl(\A[r][C_{\A}], \B_r[r][C_{\A}]\bigr)\leq d$, which represents a valid $k$-center cluster assignment. 
The reverse direction follows directly from the construction of our dynamic programming approach.
As a result, the $\kCME$ instance is feasible, if $\Dp_r \bigl(\emptyset, \emptyset, \emptyset \bigr)=1$.
Thus it remains to show that all records can be computed in a leaf-to-root order by traversing the nodes of $\mathcal{T}$. When visiting a node $t \in \mathcal{T}$, one of the following cases may arise:

\subparagraph*{\textbf{$t$ is a leaf node.}}
First assume that $X(t)$ contains only a row $r_0$, i.e. $X(t)=\{r_0\}$. In this case we have $C_t = \emptyset$, which leads to $P(t)=\bigl\{(\parti, \emptyset, \dist)|\, \parti:\{r\} \rightarrow [k] , \dist: \{r\} \rightarrow [d]\bigr\}$.
Since there are no coordinates in $\down{C_t}$ along which we can compute the Hamming distance of $\A[r_0]$ to its cluster center, by definition $\Dp_t \bigl(\parti, \emptyset, (0)\bigr) = 1$.
For every other fragment $(\parti, \emptyset, \dist) \in P(t)$ with $\dist \neq (0)$, we will have $\Dp_t\bigl(\parti, \emptyset, \dist\bigr) = 0$.
On the other hand if $X(t)=\{c_0\}$, where $c_0$ is a coordinate, then $R_t = \emptyset$ which results in $P(t)=\bigl\{(\emptyset, \cent, \emptyset)|\cent: [k] \times \{c_0\}  \rightarrow \{0, 1\}\bigr\}$.
For every $(\emptyset, \cent, \emptyset) \in P(t)$, we have $\Dp_t\bigl(\emptyset,\cent,\emptyset\bigr)=1$.

\subparagraph*{\textbf{$t$ introduces a row.}}
Let $t'$ be the child of $t$ and assume that $t$ introduces a row, meaning $X(t) = X(t') \cup \{r_0\}$.
Consequently we have $C_t = C_{t'}$ and $R_t = R_{t'} \cup \{r_0\}$.
For every $\parti':R_{t'} \rightarrow [k]$, define $\parti_i = \parti' \cup \,(r_0, i)$ with $i \in [k]$, where $\parti_i$ assigns the new row $r_0$ to cluster $i$.
For every $\dist':R_{t'} \rightarrow [d]$, let $\dist_i = \dist' \cup \, (r_0, d_0)$, where $d_0 = \hd\bigl(\A[r_0][C_t], \cent[i, C_t]\bigr)$.
If $0 \leq d_0 \leq d$ then for all $i \in [k]$, we set:
\[
\Dp_{t}\bigl(\parti_i, \cent, \dist_i\bigr) \coloneqq \Dp_{t'}\bigl(\parti', \cent, \dist'\bigr)
\]
\ifMyFlag
To see the correctness, first assume $ \Dp_{t'}\bigl(\parti', \cent, \dist'\bigr) = 1$, then by the definition there is a matrix $\B_{t'} \in \mathcal{B}_{t'}$ that confirms this. We construct a matrix $\B_t$ from $\B_{t'}$ which confirms $\Dp_{t}\bigl(\parti_i, \cent, \dist_i\bigr)=1$, and also $\dist_i$ fits it.
Observe that $\down{R}_t = \down{R}_{t'} \cup \{r_0\}$, so each $\B_t \in \mathcal{B}_t$ contains one more row than each $\B_{t'} \in \mathcal{B}_{t'}$ with the additional row labeled $r_0$.
So for every $r \in \down{R}_{t'}$, set $\B_t[r]\coloneqq\B_{t'}[r]$.
Let $\overline{C_t}= \down{C}_t\!\setminus\! C_t$.
By property \ref{tw:connected} of a nice tree decomposition, $\M[r_0][\overline{C_t}] = 0$, other wise $\overline{C_t} \cup \{r_0\}$ would appear together in a later bag, which can not happen since forgotten coordinates and rows can not be reintroduced.
Now it only remains to fill the values for row $\B_t[r_0]$ along coordinates in $C_t$ and also to adjust $\dist'$.
Let $\phi'$ be the cluster assignment with respect to $\B_{t'}$ and $\dist'$ be the distance vector that fits $\B_{t'}$. 
If for some $r \in \down{R}_{t'}$, it holds that $i = \phi'(r)$, then set $\B_t[r_0] \coloneqq \B_{t'}[r]$, otherwise for every $c \in C_t$, set $\B_t[r_0][c] \coloneqq \cent[i, c]$ and for the rest of the coordinates in $c \in \overline{C_t}$ set $\B_t[r_0][c] \coloneqq 0$. 
Note that in both cases, for every $c \in C_t$, $\B_t[r_0][c] = \cent[i, c] $: in the first case we have $\B_t[r_0][c] \coloneqq \B_{t'}[r][c] = \cent[\phi'(r), c] = \cent[i, c]$ and in the second case it follows immediately from the definition. 
So $(\parti_i, \cent)$ is a partial fragment of $\B_t$.
By assigning $r_0$ to cluster $i$ we have:
\begin{align}
	\hd \bigl(\A[r_0][\down{C}_t], \B_t[r_0][\down{C}_t]\bigr)  &= \hd \bigl(\A[r_0][\overline{C_t}], \B_t[r_0][\overline{C_t}] \bigr)                                                   + \hd \bigl(\A[r_0][C_t], \B_t[r_0][C_t] \bigr)\notag \\
	&= 0 + \hd \bigl(\A[r_0][C_t], \cent[i, C_t]\bigr) \notag = d_0 \notag
\end{align}
Since $\M[r_0][\overline{C_t}] = 0$ holds, $d_0 = \hd\bigl(\A[r_0][C_t], \cent[i, C_t]\bigr)$, represents the Hamming distance of $\A[r_0]$ to the center of cluster $i$ along coordinates in $\down{C}_t$.
As a result $\dist_i = \dist' \cup (r_0, d_0)$ fits $\B_t$, leading to $\B_t$ confirming $\Dp_{t}\bigl(\parti_i, \cent, \dist_i\bigr)=1$.\\
In the other direction, suppose that $\Dp_{t}\bigl(\parti_i, \cent, \dist_i\bigr) = 1$, then there exists a $\B_t \in \mathcal{B}_t$ that confirms it and also $\dist_i$ fits $\B_t$. 
Let $\phi$ be the cluster assignment w.r.t $\B_t$ with $(r_0, i) \in \phi$, then $\phi' = \phi \!\setminus\! (r_0, i)$ is still a cluster-assignment for $\B_t$ on rows $\down{R}_t \!\setminus\! \{r_0\} = \down{R}_{t'}$.
Also when $\dist_i$ with $(r_0, d_0) \in \dist_i$ fits $\B_t$, then $\dist'=\dist_i \!\setminus (r_0, d_0)$ fits $\B_t$ on rows $\down{R}_t \!\setminus\! \{r_0\} = \down{R}_{t'}$.
So the matrix $\B_{t'}$ obtained by removing row $r_0$ from matrix $\B_t$ confirms $\Dp_{t'}\bigl(\parti', \cent, \dist'\bigr)=1$ and $\dist'$ fits it.
\begin{figure}[h]
    \centering
    \includegraphics[width=0.4\textwidth]{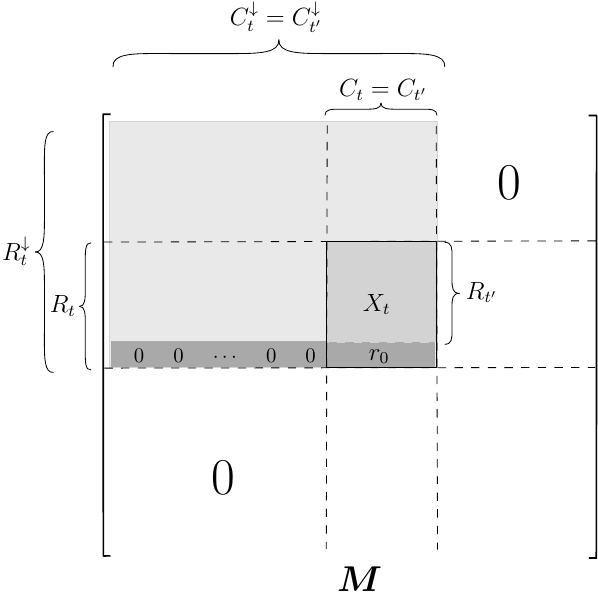} 
    \caption{Illustration of the mask matrix $\M$, when visiting node $t$ that introduces a new row $r_0$ in the nice tree decomposition. Note that $r_0$ contains zero along forgotten coordinates, i.e. $C_{\down{t}}$}
    \label{fig:introduce_row}
\end{figure}
\fi

\subparagraph*{\textbf{$t$ introduces a coordinate.}}
If $X(t) = X(t') \cup \{c_0\}$ for some coordinate $c_0$, then $R_t = R_{t'}$ and $C_t = C_{t'} \cup \{c_0\}$.
For every $\cent: [k] \times C_t  \rightarrow \{0, 1\}$, let $\cent' = \cent|_{C_{t'}}$ be the restriction of $\cent$  to all the coordinates in $C_{t'}$.
Also for each $\dist': R_{t'} \rightarrow [d]$, define $\dist$ such that for all $r \in R_{t}$ we have 
$\dist[r] = \dist'[r] + \bigl( \A[r][c_0] \oplus \cent \bigl[\parti[r], c_0 \bigr] \bigr)$, where $\oplus$ is the bit-wise XOR.
Now if dist is valid, then we set:
\[
\Dp_{t}\bigl(\parti, \cent, \dist\bigr) \coloneqq \Dp_{t'}\bigl(\parti, \cent', \dist'\bigr)
\]
\ifMyFlag
For correctness, first assume that $\Dp_{t'}\bigl(\parti, \cent', \dist'\bigr)=1$, then there is a matrix $\B_{t'} \in \mathcal{B}_{t'}$ that confirms this with fitness distance function $\dist'$.
We construct a matrix $\B_t$ from $\B_{t'}$ that confirms $\Dp_{t}\bigl(\parti, \cent, \dist\bigr)=1$ while $\dist$ fits it.
Observe that $\down{R}_t = \down{R}_{t'}$ and $\down{C}_t = \down{C}_{t'} \cup \{c_0\}$, so each $\B_t \in \mathcal{B}_t$ contains one more coordinate than each $\B_{t'} \in \mathcal{B}_{t'}$ with the additional coordinate labeled $c_0$.
So for every $r \in \down{R}_{t'}$ and $c \in \down{C}_{t'}$, set $\B_t[v][c]\coloneqq\B_{t'}[v][c]$.
Now it only remains to fill the value for the coordinate $c_0$ along all rows $r \in \down{R}_{t}$ and also to adjust $\dist$.
Let $\overline{R_t} = \down{R}_{t} \setminus R_t$, then by property \ref{tw:connected} of a nice tree decomposition, for all of the forgotten rows $r_f \in \overline{R_t}$, it holds that $\M[r_f][c_0]=0$.
So we can simply put $\B_t[r_f][c_0]\coloneqq 0$, for each forgotten row $r_f \in \overline{R_t}$.
Let $\phi'$ be the cluster assignment with respect to $\B_{t'}$ (i.e. $\parti=\phi'|_{R_{t'}}$).
For $r \in R_{t'}$, set $\B_t[r][c_0] \coloneqq \cent[\phi'(r), c_0]$.
Note that, for every $c \in C_t$ and $r \in \down{R}_{t'}$, it holds that $\B_t[r][c] = \cent[\phi'(r), c] $: in the first case we have $\B_t[r][c] := \B_{t'}[r][c] = \cent'[\phi'(r), c] = \cent[\phi'(r), c]$ and in the second case it follows immediately from the definition. 
As a result $(\parti, \cent)$ with the same cluster assignment $\phi'$ is a partial fragment of $\B_t$.
Let $\dist'=(d'_1,\dotso, d'_{n_{R_{t'}}})$ be the distance vector that fits $\B_{t'}$. 
Since For $r_f \in \overline{R_t}$ it holds that $\M[r_f][c_0]=0$, only the distances between rows in the bag i.e. $R_t$ and their assigned cluster centers, as defined by $\phi'$, are updated. 
Hence, for $r \in R_t$
\begin{align}
	\hd\bigl(\A[r][\down{C}_t], \B_t[r][\down{C}_t]\bigr)  &= \hd\bigl(\A[r][\down{C}_t\!\setminus\! c_0], \B_t[r][\down{C}_t\!\setminus\! c_0]\bigr) 
                                                + \hd\bigl(\A[r][c_0], \B_t[r][c_0]\bigr) \notag \\ 
	                                        &= \hd\bigl(\A[r][\down{C}_{t'}], \B_t[r][\down{C}_{t'}]\bigr)
                                                + \hd\bigl(\A[r][c_0], \B_t[r][c_0]\bigr) \notag \\ 
	                                        &= \dist'[r] +\bigl(\A[r][c_0] \oplus \cent\bigl[\parti[r], c_0 \bigr] \bigr) \notag \\
                                                &= \dist[r], \notag
\end{align}
as a result $\dist[r]$ for all $r \in R_t$, fits $\B_t$ leading to $\B_t$ confirming $\Dp_{t}\bigl(\parti, \cent, \dist\bigr)=1$. \\
On the other hand, suppose that $\Dp_{t}\bigl(\parti, \cent, \dist\bigr) = 1$, then there exists a $\B_t \in \mathcal{B}_t$ that confirms it and also a $\dist$ which fits $\B_t$. 
Let $\phi$ be the cluster assignment w.r.t $\B_t$, that is $\parti = \phi|_{R_t}$.
Then removing coordinate $c_0$ from $\B_{t}$, yields a matrix $\B_{t'}$ with the same cluster assignment $\phi$ where $\B_{t'}[r][\down{C}_{t'}] = \B_{t}[r][\down{C}_t \!\setminus\! c_0] = \B_{t}[r][\down{C}_{t'}]$ for every $r \in \down{R}_{t'}$.
So for every $c' \in C_{t'}$ and $r \in R_{t'}$, it holds that:
$$\B_{t'}[r][c'] = \B_{t}[r][c'] = \cent[\phi(r), c'] =  \cent'[\phi(r), c']$$
Thus $(\parti, \cent')$ is a partial fragment of $\B_{t'}$ at $t'$.
Moreover, the distance between each $r \in R_{t'}$  to its respective cluster center, as determined by $(\parti, \cent')$ is calculated as $\dist[r] - \bigl( \A[r][c_0] \oplus \cent[\parti[r], c_0 ] \bigr)$.
So $\dist'$ fits $\B_{t'}$ leading to $\Dp_{t'}\bigl(\parti, \cent', \dist'\bigr)=1$ being confirmed by $(\parti, \cent', \dist')$.\\
\fi

\subparagraph*{\textbf{$t$ forgets a row.}}
Let $t'$ be the child of $t$, and assume that $t$ forgets a row, that is, $X(t) = X(t') \!\setminus\! \{r_0\}$. 
For each $\parti:R_t \rightarrow [k]$ and $\dist: R_{t'} \rightarrow [d]$, respectively define sets $Y_{part}= \bigl\{\parti'_i: R_{t'} \rightarrow [k]: \, |\,\parti'_i = \parti \cup (r_0, i), \, \forall i \in [k] \bigr\}$ and $D_{\dist} = \bigl\{\dist'_j: R_{t'} \rightarrow [d] \; \big| \dist'_j = \dist \cup \, (r_0, j), \, \forall j \in [d] \bigr\} $ to be their respective extensions at node $t'$.
For a fragment $(\parti, \cent, \dist) \in P(t)$, we set:
\[
\Dp_t \bigl(\parti, \cent, \dist\bigr) \coloneqq \bigvee_{\substack{\forall \parti'_i  \in \, Y_{\parti} , \\ \forall \dist'_j \in \, D_{\dist}}} {\Dp_{t'} \bigl(\parti'_i, \cent, \dist'_j\bigr)} 
\]
\ifMyFlag
To start with correctness, assume that there exists at least one fragment $(\parti'_i, \cent, \dist'_j) \in P(t')$ such that $\Dp_{t'} \bigl(\parti'_i, \cent, \dist'_j\bigr) = 1$.
This implies that there is a matrix $\B_{t'} \in \mathcal{B}_{t'}$ confirming it. 
Since $\down{R}_t = \down{R}_{t'}$ and $\down{C}_t = \down{C}_{t'}$, keep in mind that $\mathcal{B}_t = \mathcal{B}_{t'}$.
Therefor, for every $r \in \down{R}_t$ and $c \in \down{C}_t$, we set $\B_t[r][c] \coloneqq \B_{t'}[r][c]$.
Let $\phi'$ be the cluster assignment corresponding to $\B_{t'}$ where $\parti'_i= \phi' \big|_{R_{t'}} $, then, $(\parti, \cent)$ with $\parti = \parti'_i\!\setminus (r_0, i)$ becomes a partial fragment of $\B_t$ at $t$.
Regarding the distances, let $\dist = \dist'_j \setminus (r_0, j)$.
Since $\B_{t'}$ confirms $\Dp_{t'} \bigl(\parti', \cent, \dist'_j\bigr) = 1$, by definition \ref{dp:b} of our dynamic programming record, for all $r \in R_{t'}$ we have 
$$\hd \bigl(\A[r][\down{C}_{t'}], \B_{t'}[r][\down{C}_{t'}]\bigr) = \dist'_j[r].$$
Which ensures that  for all $r \in R_{t}$: 
$$\hd \bigl(\A[r][\down{C}_{t}], \B_{t}[r][\down{C}_{t} ]\bigr) = \dist[r] $$
and
\begin{equation}
   \hd \bigl(\A[r_0][\down{C}_{t}], \B_{t}[r_0][\down{C}_{t}] \bigr) \leq d
   \label{eq:forgetrow}
\end{equation}
Similarly, by definition \ref{dp:c} of our dynamic programming record, for all $r \in (\down{R}_{t'} \setminus R_{t'})$, we have 
$$\hd \bigl(\A[r][\down{C}_{t'}], \B_{t'}[r][\down{C}_{t'}]\bigr) \leq d$$
which for all $r \in (\down{R}_{t} \setminus R_t)$ along with \cref{eq:forgetrow} translates to , 
$$\hd \bigl( \A[r][\down{C}_{t}], \B_{t}[r][\down{C}_{t}]\bigr) \leq d.$$
As a result $\B_t$ confirms that $\Dp_t \bigl(\parti, \cent, \dist \bigr)=1$.\\
Conversely, if for a fragment $(\parti, \cent, \dist) \in P(t)$ it  holds that $\Dp_t \bigl(\parti, \cent, \dist\bigr) = 1$ , then there is a matrix $\B_{t} \in \mathcal{B}_{t}$ that confirms this.
Let $\phi$ denote the cluster assignment w.r.t $\B_t$.
Then $(\parti', \cent )$ with $\parti'_i= \bigl(\parti \cup \, (r_0, \phi(r_0))\bigr)$ forms a partial fragment of $\B_{t'}$ at $t'$.
Let $d_0 = d_H \bigl(\A[r_0][\down{C}_t], \B_t[r_0][\down{C}_t]\bigr)$,  by property \ref{dp:c} we have $d_0 \leq d$ and: 
$$d_H \bigl(\A[r_0][\down{C}_t], \B_t[r_0][\down{C}_t]\bigr) \leq d,$$
so define $\dist' = \dist \cup \, (r_0, d_0) $, then for all $r' \in R_{t'}$ it holds that:
$$d_H \bigl(\A[r'][\down{C}_t], \B_t[r'][\down{C}_t]\bigr) = \dist'[r']$$
and for all $r' \in (\down{R}_{t'} \setminus R_{t'})$:
$$d_H \bigl(\A[r'][\down{C}_t], \B_t[r'][\down{C}_t]\bigr) \leq d$$
As a result, $\B_t$ confirms that $\Dp_t (\parti'_i,\cent, \dist')=1$.\\

\subparagraph*{\textbf{$t$ forgets a coordinate.}}
Assume that $t$ forgets a coordinate, that is, $X(t) = X(t') \!\setminus\! \{c_0\}$. 
For each $\cent:[k] \times C_t \rightarrow \{0,1\}$, let the $C_{\cent} = \bigl\{\cent': [k] \times C_{t'} \rightarrow \{0,1\} \, \big| \, \cent = \cent'|_{C_{t}}\bigr\}$ to be the set of its respective extensions at node $t'$.
For a fragment $(\parti, \cent, \dist) \in P(t)$, we set:
\[
\Dp_{t} \bigl(\parti, \cent, \dist \bigr) \coloneqq \bigvee_{\substack{\cent' \in \, C_{\cent}}} \Dp_{t'} \bigl(\parti, \cent', \dist \bigr)
\]
\ifMyFlag
For correctness, first assume that $\Dp_{t'}\bigl(\parti, \cent', \dist \bigr) = 1$, for a fragment $(\parti, \cent', \dist) \in P(t')$ with $\B_{t'} \in \mathcal{B}_{t'}$ confirming it. 
Again note that $\mathcal{B}_{t}=\mathcal{B}_{t'}$.
For each $r \in \down{R}_t$ and $c \in \down{C}_t$, we assign $\B_t[r][c] \coloneqq \B_{t'}[r][c]$.
Let $\phi'$ represent the cluster assignment associated with $\B_{t'}$, then by \ref{dp:a}, for every $r \in \down{R}_t$ and $c \in C_t$, we have $\B_t[r][c] = \cent'[\phi'(r), c]$.
Thus $(\parti, \cent)$ with $\cent = \cent'|_{C_t}$, forms a partial fragment for $\B_t$ at node $t$.
Furthermore, based on the fact that $R_t = R_{t'}$, $\down{R}_t = \down{R}_{t'}$, and $\down{C}_t = \down{C}_{t'}$, it follows from definitions \ref{dp:b} and \ref{dp:c} that $\hd \bigl(\A[r][\down{C}_{t}], \B_{t}[r][\down{C}_{t}] \bigr) = \dist[r]$, for all $r \in R_{t}$.
Additionally, for all $r \in  \down{R}_{t} \!\setminus\! R_{t}$, $\hd \bigl(\A[r][\down{C}_{t}], \B_{t}[r][\down{C}_{t}]\bigr) \leq d$ .
Consequently $\B_t$ confirms that $\Dp \bigl(\parti, \cent, \dist \bigr)=1$.\\
Alternatively if for a fragment $(\parti, \cent, \dist) \in P(t)$, it holds that $\Dp_t \bigl(\parti, \cent, \dist \bigr) = 1$ , then there is a matrix $\B_{t} \in \mathcal{B}_{t}$ that confirms this. 
Again for each $r \in \down{R}_t$ and $c \in \down{C}_t$, set $\B_{t'}[r][c] \coloneqq \B_t[r][c]$.
Let $\hat{C}= \bigcup_{r \in R_t} \bigl((c_0, \phi(r)), \B_t[\phi(r)][c_0]\bigr)$ and set $\cent' \coloneqq \cent \cup \, \hat{C} $, then $(\parti, \cent')$ is a partial fragment of $\B_{t'}$ at $t'$.
It is easy to conclude that \ref{dp:b} and \ref{dp:c} directly hold for $\B_{t'}$.
Consequently, $\B_{t'}$ confirms that $\Dp_t \bigl(\parti ,\cent', \dist \big)=1$.\\

\subparagraph*{\textbf{$t$ is a join node.}}
Assume that $t_1$ and $t_2$ are the children of $t$, meaning, $X(t) = X(t_1) = X(t_2)$. 
Define $Y_t = \bigl\{(\parti, \cent)|\parti: R_t \rightarrow [k]\, ,\,  \cent: [k] \times C_t \rightarrow \{0,1\}\bigr\}$ be the set of all partial fragments at $t$.
Also, let $\Delta:Y_t \rightarrow [d]$ be the function such that for every $(\parti, \cent) \in Y_t$, it holds that $\Delta (\parti, \cent) = \bigl(\delta_1, \delta_2, \dotso, \delta_{n_{R_t}}\bigr)$, where $\delta_i = \hd \bigl(\A[r_i][C_t], \cent \bigl[\parti[r_i], C_t\bigr] \bigr)$ for every $r_i \in R_t$. Then for each $(\parti, \cent, \dist) \in P(t)$, we set:
\[
\Dp_t \bigl (\parti, \cent, \dist\bigr) \coloneqq \Dp_{t_1} \bigl(\parti, \cent, \dist_1\bigr) \; \bigwedge \; \Dp_{t_2} \bigl(\parti, \cent, \dist_2\bigr) 
\]
where $\dist  = \dist_1 + \dist_2 - \Delta(\parti, \cent)$ with $\dist_i: R_t \rightarrow [d]$ for $i \in \{1, 2\}$.\\
For correctness, first assume $\Dp_{t_i} \bigl (\parti, \cent, \dist_i\bigr) =1$ and that $\B_{t_i}$ confirms it, for $i \in \{0, 1\}$.
We obtain a matrix $B_t \in \mathcal{B}_t$ that confirms $\Dp_t \bigl(\parti, \cent, \dist\bigr) = 1$ with $\dist$ that fits $B_t$.
Note that $\down{R}_t =\down{R}_{t_1} \cup \down{R}_{t_2} $ and $\down{C}_t =\down{C}_{t_1} \cup \down{C}_{t_2}$.
Let $\phi_i$ be the cluster assignment corresponding to $\B_{t_i}$, then $\parti = \phi_1 |_{R_{t_1}} = \phi_1 |_{R_{t_2}}$.
Consequently, we have $\B_{t_1}[r][c]=\cent \bigl[\parti[r], c \bigr]=\B_{t_2}[r][c]$, for each $r \in R_t$ and $c \in C_t$.
Also remember that, for every $r \in (\down{R}_{t_i}\setminus R_{t_i})$ and $c \in (\down{C}_{t}\setminus C_{t_i})$, the mask matrix $\M[r][c] = 0$, since otherwise the forgotten row $r$ and the forgotten coordinate $c$ would need to appear later in a bag, which would contradict property \ref{tw:connected} of the nice tree decomposition.
Thus, the entries of $\B_{t}$ can be filled as follows:
\begin{itemize}
	\item  For every $r \in \down{R}_{t_i}$ and $c \in \down{C}_{t_i}$, set $\B_t[r][c]\coloneqq\B_{t_i}[r][c]$
	\item  For every $r \in (\down{R}_{t_i}\setminus R_{t_i})$ and $c \in (\down{C}_{t} \setminus \down{C}_{t_i})$, set $\B_t[r][c]\coloneqq 0$,
\end{itemize}
As a result, $(\parti, \cent)$ with cluster assignment $\phi = (\phi_1 \cup \phi_2)$, will be a partial fragment for $\B_t$.
Based on the assignment of $\B_t$, only the distances between the rows within the current bag to their corresponding cluster centers will change while for the remaining rows, the distances stay unchanged.
Therefore, for all $r \in (\down{R}_t \setminus R_t)$, it holds that $\hd \bigl(\A[r][\down{C}_t], \B_t[r][\down{C}_t]\bigr) \leq d$.
Let $\overline{C_{t_i}}$ represent the forgotten coordinates at node $t_i$. 
Then, for the rows inside the current bag, $r \in R_t$, the distance to their cluster centers over $\down{C}_t$ can be written as follows: 
\begin{align*}
	\hd \bigl(\A[r][\down{C}_t], \B_t[r][\down{C}_t] \bigr) &= \sum_{i=1}^{2}\hd \bigl(\A[r][\overline{C_{t_i}}], \B_t[r][\overline{C_{t_i}}] \bigr)
        + \hd \bigl(\A[r][C_t], \B_t[r][C_t] \bigr) \\ 
	&= \dist_1[r] + \dist_2[r]
        - \hd\bigl(\A[r][C_t], \B_t[r][C_t] \bigr) \\ 
	&= \dist_1[r] + \dist_2[r] - \delta_r, 
\end{align*}
since for each $r \in R_t$, we have 
$$\dist_i[r] = \hd \bigl(r[\overline{C_{t_i}}] + \hd\bigl(r[C_t], \B_t[r, C_t] \bigr),$$
it follows that $\dist = \dist_1 + \dist_2 - \Delta$, fits $\B_t$.
As a result, $\B_t$ confirms $\Dp_t \bigl (\parti, \cent, \dist\bigr) = 1$.
Now suppose there is a matrix $\B_t \in \mathcal{B}_t$ with cluster assignment $\phi$, confirming that $\Dp_t \bigl (\parti, \cent, \dist\bigr)=1$ for a fragment $(\parti, \cent, \dist) \in P(t)$.
We show there is a partial solution $\B_{t_i}$ at node $t_i$ with distance vector $\dist_i$ such that $\Dp_{t_i} \bigl (\parti, \cent, \dist_i\bigr)=1$, for $i \in \{1, 2\}$.
With an abuse of notation, let $\phi|_{X_{t_i}}$, denote the restriction of $\phi$ to rows and coordinates within the bag of node $t_i$.
Now assign $\B_{t_i} = \B_t |_{\down{R}_{t_i} \times \down{C}_{t_i}}$.
Since $\down{X}_{t_i} \subset \down{X}_t$, according to property \ref{dp:a} at node $t$, $\phi_i = \phi|_{X_{t_i}}$ is a valid cluster assignment for $\B_{t_i}$. 
Consequently, $(\parti, \cent)$ is a partial fragment for both $\B_1$ and $\B_2$.
Moreover, according to property \ref{dp:b}at node $t$ and the fact that $\down{C}_{t_i} \subset \down{C}_t$, for every $r \in \down{R}_{t_i} \setminus R_{t_i}$, it holds that $\hd \bigl(r[\down{C}_{t_i}], \B_{t_i}[r, \down{C}_{t_i}]\bigr) \leq d$.
On the other hand, for $r \in R_t = R_{t_i}$, by property \ref{dp:c} of $\dist$, we know that $\sum_{i=1}^{2}\hd \bigl(\A[r][\overline{C_{t_i}}], \B_t[r][\overline{C_{t_i}}] \bigr) + \hd \bigl(\A[r][C_t], \B_t[r][C_t] \bigr) = \dist[r]$, which translates to $\hd \bigl(\A[r][\down{C}_{t_i}], \B_t[r][\down{C}_{t_i}] \bigr) \bigr) \leq \dist[r]$ for each $i \in \{1,2\}$.
So there is $\dist_i: R_{t_i} \rightarrow [d]$, such that for $r \in R_{t_i}$ it holds that $\hd \bigl(\A[r][\down{C}_{t_i}], \B_t[r][\down{C}_{t_i}] \bigr) \bigr) = \dist_i[r]$.
As a result, there was already a fragment $(\parti, \cent, \dist_i) \in P(t_i)$ for which $\Dp_{t_i}\bigl(\parti, \cent, \dist_i\bigr) =1 $ and a partial solution $\B_{t_i}$ to confirm it.\\
\fi


The nice tree decomposition can be obtained in $2^{\bo{\tw(G_{\M})}}(n+m)$.
Furthermore, when filling the records $D_t$ in our dynamic programming, at each step we need to evaluate $k^{\tw(G_{\M})}=2^{k\cdot {\tw(G_{\M})}}$ possible mappings for $\parti$, $2^{k\cdot\tw(G_{\M})}$ mappings for $\cent$ and $d^{\tw(G_{\M})}$ different vectors for $\dist$.
iven that the number of nodes in the nice tree decomposition is linear in $n$, the overall running time of the algorithm is bounded above by $2^{\bo{k \cdot \tw(G_{\M})}}\cdot d^{\bo{\tw(G_{\M})}} \poly(n)$.
 
\end{proof}

    \section{Fracture number}
     \label{sec:fr_fpt}
    In this section, we present a fixed-parameter algorithm for the $\kCME$, where the parameter is the fracture number of the incidence graph $G_{\M}$.
To provide context, we first define the key concepts of \textit{fracture modulator} and \textit{fracture number}.
Following this, we exploit the relevant results from \cref{thm:vc} and \cref{thm:tw}, which lead to the proof of \cref{thm:fr}.
\thmfr*

\begin{figure}[h]
    \centering
    \includegraphics[width=0.4\linewidth]{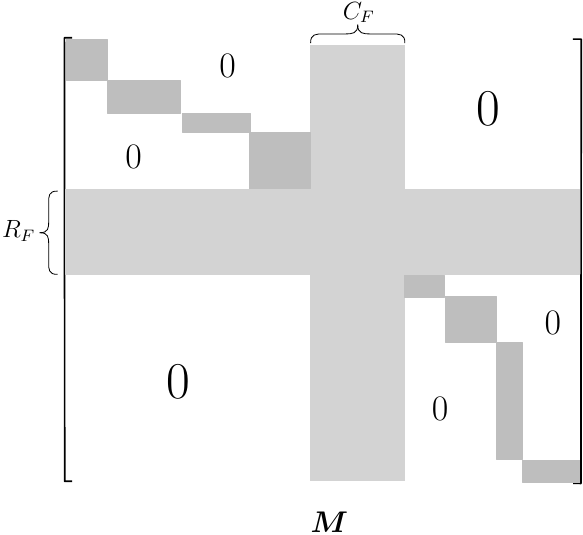}
    \caption{The rows $R_F$ and columns $C_F$ of the fracture modulator are in light gray, the ``blocks'' induced by the connected components of $G_{\M} \setminus F$ are in dark gray. All entries of $\M$ outside of the gray areas are $0$. }
    \label{fig:fracture_matrix}
\end{figure}

For a graph $G = (V, E)$, a \textit{fracture modulator} is a subset of vertices $F \subseteq V$ such that, after removing the vertices in $F$, each remaining connected component contains at most $|F|$ vertices; the size of the smallest fracture modulator is denoted by $\fr(G)$.
Consider the incidence graph $G_{\M}$ corresponding to the mask matrix $\M$, and let $F$ represent the fracture modulator of $G_{\M}$ with the smallest cardinality. We note that $F$ can be computed in time $\bo{(\fr(G)  + 1)^{\fr(G)} nm}$ by the algorithm of \citep{DvorakEGKO21}.
Let $R_{F} = {(R_{\M} \cap F)}$ and $C_{F} = { (C_{\M} \cap F)}$ represent the row and column vertices of $G_{\M}$ that belong to the fracture modulator $F$, respectively.
Additionally, define ${\overline{R_F}} = R_{\M} \setminus R_F$ and ${\overline{C_F}} = C_{\M} \setminus C_F$.
We refer to the rows in $R_F$ as \emph{long} rows while the rows corresponding to $\overline{R_F}$ are called \emph{short} rows. See \Cref{fig:fracture_matrix} for an illustration of the structure of the instance. Now, the algorithm considers two cases.
\vspace{-.4cm}
\subparagraph*{\textbf{$\bm{2\fr(G_{\M}) \le d}$.}} In this case, each of the short rows has distance at most $d$ to any fixed vector. This holds since for each row $i \in \overline{R_F}$, $\M[i][j] = 1$ only for $j \in C_F$ or $j$ in the same connected component of $G_{\M} \setminus F$ as $i$; this is at most $2\fr(G_{\M}) \le d$ entries, and all other entries in the row are missing. Therefore, the short rows are essentially irrelevant for the solution, as they can be assigned to any cluster with any center vector. We run the algorithm of \Cref{thm:vc} on the instance restricted to the long rows $R_F$, and complement the resulting solution with an arbitrary assignment of the short rows to the $k$ clusters. Since the vertex cover of the restricted instance is at most $|R_F| \le \fr(G_{\M})$, the running time bound holds as desired.
\vspace{-.4cm}
\subparagraph*{\textbf{$\bm{2\fr(G_{\M}) > d}$.}} Here, we use the algorithm of \Cref{thm:tw} to solve the given instance. We observe that a tree decomposition of width at most $2\fr(G_{\M})$ can be constructed for $G_{\M}$ in a straightforward fashion: create a bag for each connected component of $G_{\M} \setminus F$ containing all vertices of this component together with $F$, and arrange these bags on a path in arbitrary order. Since $d < 2\fr(G_{\M})$, the running time of \Cref{thm:tw} gives the desired bound.
    \section{Conclusion}
     \label{sec:conclusion}




We have investigated the algorithmic complexity of \kCME in the setting where the missing entries are sparse and exhibit certain graph-theoretic structure. We have shown that the problem is FPT when parameterized by $\vc + k$, $\fr + k$ and $\tw + k + d$, where $\vc$ is the vertex cover number, $\fr$ is the fracture number, and $\tw$ is the treewidth of the incidence graph.
In fact, it is not hard to get rid of $k$ in the parameter; for example, in the enumeration of partial center assignments in the algorithm of \Cref{thm:vc} it can be assumed that $k \le 2^{\vc}$, as multiple centers that have the same partial assignment are redundant.
However, this would increase the running time as a function of the parameter to doubly exponential, therefore we state the upper bounds with explicit dependence on $k$. It is, on the other hand, an interesting open question, whether $d$ in the parameter is necessary for the algorithm parameterized by $\tw$. As shown by \citep{GanianHKOS22}, this is not necessary for \kMME, since the problem admits an FPT algorithm when parameterized by treewidth alone. Yet, it does not seem that the dynamic programming approaches used in their work and in our work, can be improved to avoid the factor of $d^{\bo{\tw}}$ in the case of \kCME. Therefore, it is natural to ask whether it can be shown that \kCME is W[1]-hard in this parameterization.

Another intriguing open question is the tightness of our algorithm in the parameterization by the vertex cover number (and fracture number). On the one hand, the running time we show is $2^{\vc^{2 + o(1)}} \cdot \poly(nm)$ (for values of $k$ in $\bo{vc}$), exceeding the ``natural'' single-exponential in $\vc$ time, and we are not aware of a matching lower bound that is based on standard complexity assumptions.
On the other hand, we show that improving this running time improves also the best-known running time for \ILP when parameterized by the number of constraints and \CS parameterized by the number of strings. The latter are major open questions; \citep{RohwedderW2024} show also that several other open problems are equivalent to these.
On the positive side, our algorithm in fact reduces \kCME to just $2^{\vc}$ instances of \ILP. Since practical ILP solvers are quite efficient, coupling our reduction with an ILP solver is likely to result in the running time that is much more efficient than prescribed by the upper bound of \Cref{thm:vc}. 

    \newpage
    \bibliographystyle{plainnat}
    \bibliography{bibliography}

\end{document}